\documentclass[12pt]{article}
\usepackage{amsmath}
\usepackage{enumerate}

\author{Michael Schweinberger \and Sergii Babkin \and Katherine B.\ Ensor\thanks{Address: Department of Statistics, Rice University, 6100 Main St, Houston, Texas 77005. Email: michael.schweinberger@rice.edu.}}

\newcommand{\ghost}[0]{}
\newcommand{\ghoster}[0]{}

\usepackage{url} 
\usepackage[letterpaper, portrait, margin=1in]{geometry}

\usepackage[colon]{natbib}
\bibpunct{(}{)}{,}{a}{}{,}

\newcounter{proof}
\newenvironment{proof}[1][]{\refstepcounter{proof}\par\medskip\noindent%
	\textit{Proof.} \rmfamily}{\medskip}

\newcounter{assumption}
\newenvironment{assumption}[1][]{\refstepcounter{assumption}\par\medskip\noindent%
\textit{Assumption~\theassumption #1}. \rmfamily}{\medskip}

\newcounter{lemma}
\newenvironment{lemma}[1][]{\refstepcounter{lemma}\par\medskip\noindent%
\textbf{Lemma~\thelemma #1}. \rmfamily}{\medskip}

\newcounter{theorem}
\newenvironment{theorem}[1][]{\refstepcounter{theorem}\par\medskip\indent%
\textbf{Theorem~\thetheorem #1}. \rmfamily}{\medskip}

\usepackage{times}
\usepackage{bm}
\usepackage{multirow}
\usepackage[plain,noend]{algorithm2e}

\usepackage[mathscr]{eucal}

\usepackage{bbm,amsfonts,graphicx}

\newcommand{\mQ}{\mathbb{Q}(\bbeta^\star, \bSigma)}

\newcommand{\mA}{\mathscr{A}}
\newcommand{\mB}{\mathscr{B}}
\newcommand{\mC}{\mathscr{C}}

\newcommand{\mR}{\mathbb{R}}

\newcommand{\supp}{\mathbb{S}}
\newcommand{\mG}{\mathscr{G}}

\newcommand{\mN}{\mathscr{N}}
\newcommand{\s}{\vspace{0.25cm}}

\newcommand{\dsum}{\displaystyle\sum\limits}
\newcommand{\mbP}{\mathbb{P}}

\newcommand{\wh}{\widehat}
\newcommand{\bX}{\bm{X}}

\newcommand{\bA}{\bm{A}}

\newcommand{\bD}{\bm{E}}
\newcommand{\bI}{\bm{I}}
\newcommand{\bee}{\bm{e}}
\newcommand{\bSigma}{\bm{\Sigma}}

\newcommand{\bE}{\bm{E}}
\newcommand{\bXX}{\bm{\mathcal{X}}}
\newcommand{\bYY}{\bm{\mathcal{Y}}}
\newcommand{\bbeta}{\bm{\beta}}

\newcommand{\bgamma}{\bm{\gamma}}
\newcommand{\bGamma}{\bm{\Gamma}}
\newcommand{\bb}{\bm{b}}

\newcommand{\bv}{\widehat{\bm{v}}} %
\newcommand{\bvs}{\widehat{\bm{v}}_{\supp[0,\rho-\delta]}} %
\newcommand{\bvc}{\widehat{\bm{v}}_{\overline\supp[0,\rho-\delta]}} %

\newcounter{com}
\newenvironment{com}[1][]{\refstepcounter{com}\par\medskip\indent%
\textit{Remark~\thecom #1}. \rmfamily}{\medskip}

\newcommand{\mS}{\mathscr{S}}
\newcommand{\wwm}{\widehat{\rho}\hspace{.125cm}}
\newcommand{\wm}{\widehat{\rho}\hspace{.125cm}}

\newcommand{\beno}{\begin{equation}\begin{array}{llllllllll}\nonumber}
\newcommand{\be}{\begin{equation}\begin{array}{llllllllll}}
\newcommand{\ee}{\end{array}\end{equation}}
\newcommand{\norm}[1]{\lVert#1\rVert}

\newcommand{\mbR}{\mathbb{R}}

\newcommand{\mbZ}{\mathbb{Z}}
\newcommand{\mZ}{\mathbb{Z}}

\newcommand{\mbS}{\mathbb{S}}

\newcommand{\hide}[1]{}
\newcommand{\alert}{\textcolor{red}}
\DeclareMathOperator*{\argmin}{arg\,min}

\newcommand{\vv}{\mbox{vec}}

\title{High-Dimensional Multivariate Time Series With Additional Structure}

\date{}

\begin{document}

\maketitle 

\begin{abstract}
High-dimensional multivariate time series are challenging due to the dependent and high-dimensional nature of the data,
but in many applications there is additional structure that can be exploited to reduce computing time along with statistical error.
We consider high-dimensional vector autoregressive processes with spatial structure, 
a simple and common form of additional structure.
We propose novel high-dimensional methods that take advantage of such structure without making model assumptions about how distance affects dependence.
We provide non-asymptotic bounds on the statistical error of parameter estimators in high-dimensional settings and show that the proposed approach reduces the statistical error.
An application to air pollution in the U.S.A.\ demonstrates that the estimation approach reduces both computing time and prediction error and gives rise to results that are meaningful from a scientific point of view, 
in contrast to high-dimensional methods that ignore spatial structure.
In practice,
these high-dimensional methods can be used to decompose high-dimensional multivariate time series into lower-dimensional multivariate time series that can be studied by other methods in more depth. 

\end{abstract}

\section{Introduction}
\label{sec:introduction}

Multivariate time series \citep[e.g.,][]{Lu07,wilson2015models} arise in a wide range of applications,
from finance to studies of air pollution and ecological studies \citep[e.g.,][]{EnRa13,Ho13,ChWa15}.
The age of computing has made it possible to collect data sets with large numbers of time series,
where the number of parameters may exceed the number of observations.
A common approach to dealing with high-dimensional data is to endow models with additional structure in the form of sparsity \citep[e.g.,][]{BuGe11}.
In the case of high-dimensional multivariate time series,
an additional challenge is the complex dependence within and between time series.
Some consistency results on model estimation and selection of high-dimensional vector-autoregressive processes were obtained by \citet{SoBi11},
though under strong assumptions.
\citet{LoWa11} and \citet{BaMi13} developed powerful concentration inequalities that enabled them to establish consistency under weaker conditions and prove that these conditions hold with high probability.
In particular,
\citet{BaMi13} established consistency of $\ell_1$-penalized least squares and maximum likelihood estimators of the autoregressive coefficients of high-dimensional vector autoregressive processes and related the estimation and prediction error to the complex dependence structure of vector autoregressive processes.
Other estimation approaches,
including Bayesian approaches, 
are discussed by \citet{DaZa12} and \citet{NgKa14}.

\ghoster{
\begin{figure}[t]
\begin{center}
\includegraphics[scale=.4]{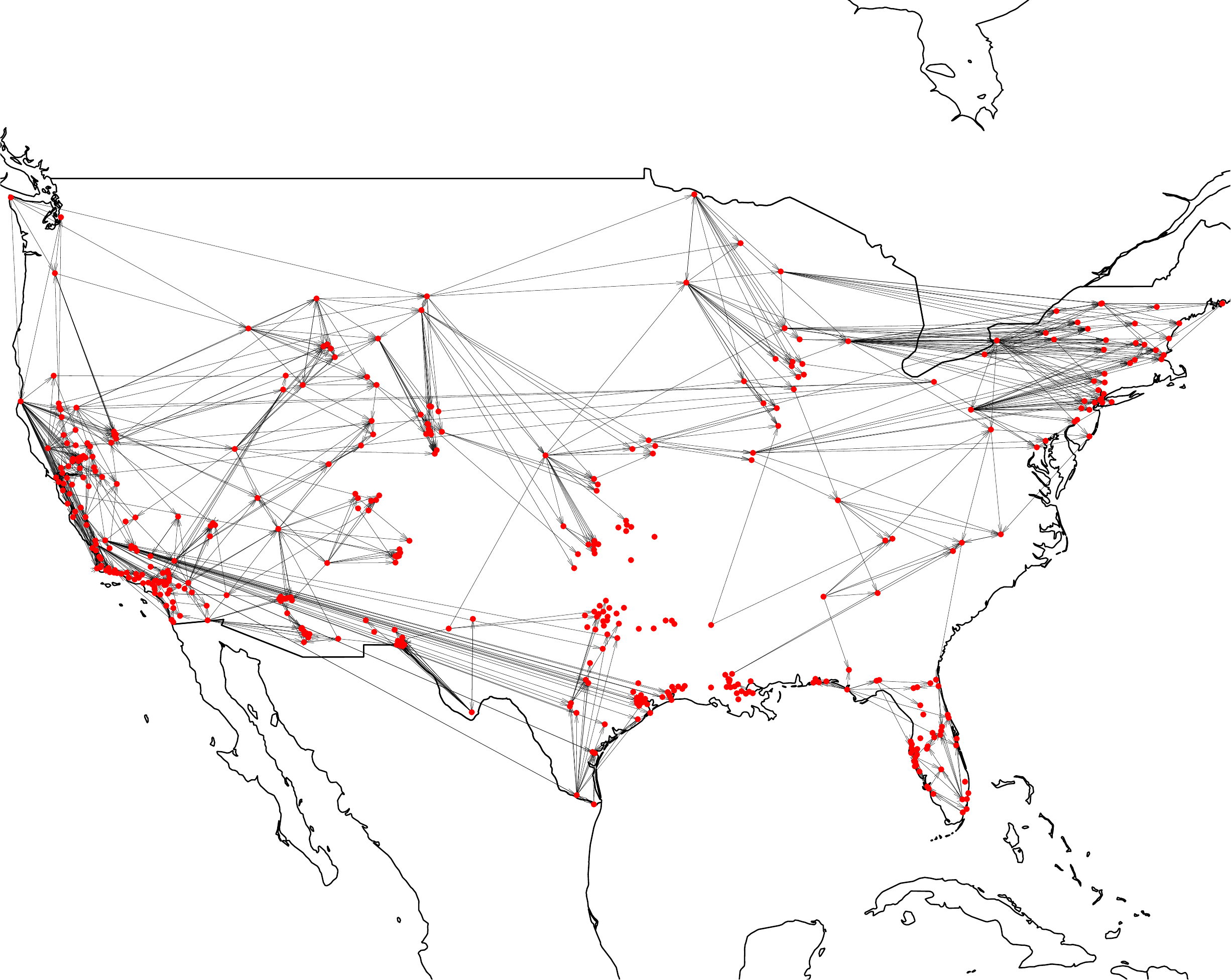} 
\caption{
\label{comparison1}
Air pollution in the U.S.A.: autoregressive coefficients estimated by the $\ell_1$-penalized least squares method from daily measurements of Ozone. 
Monitors are connected by edges if the estimates of the corresponding autoregressive coefficients are non-zero. 
The long-distance edges contradict scientific evidence \citep[see, e.g.,][]{Raetal97}.}
\end{center}
\end{figure}
}

We consider high-dimensional vector autoregressive processes with $p \gg N$ parameters,
where $p$ is the number of parameters and $N$ is the number of observations.
While high-dimensional vector autoregressive processes are challenging due to the dependent and high-dimensional nature of the data,
in many applications there is additional structure that can be exploited to reduce computing time along with statistical error.
Examples are studies of air pollution and ecological studies,
where spatial structure can help reduce computing time and statistical error.
If such structure is ignored,
high-dimensional methods can give rise to results that contradict science.
An example are daily measurements of Ozone recorded by monitors across the U.S.A.\ as described in Section \ref{sec:application}.
Figure \ref{comparison1} shows the non-zero pattern of autoregressive coefficients estimated by the $\ell_1$-penalized least squares method described in Section \ref{sec:step1}.
The figure suggests that today's Ozone levels on the East Coast can directly affect tomorrow's Ozone levels on the West Coast.
Such results contradict science,
because Ozone cannot travel long distances \citep[see, e.g.,][]{Raetal97}.

We introduce novel methods and theory that take advantage of additional structure in the form of space with a view to reducing computing time along with statistical error,
without making model assumptions about how the distance between the components of the vector autoregressive process affects the dependence between the components. 
We provide non-asymptotic bounds on the statistical error of parameter estimators in high-dimensional settings and show that the proposed approach reduces the statistical error.
An application to air pollution recorded by 444 monitors across the U.S.A.\ with $N =$ 1,826 observations and $p =$ 197,136 parameters demonstrates that the proposed methods reduce both computing time and prediction error compared with existing high-dimensional methods and gives rise to results that are meaningful from a scientific point of view, 
in contrast to high-dimensional methods that ignore the spatial structure.
In practice,
these high-dimensional methods can be used to decompose high-dimensional multivariate time series into lower-dimensional multivariate time series that can be studied by other methods in more depth. 

The paper is structured as follows.
We introduce vector autoregressive processes in Section \ref{sec:models}.
Methods and theory are described in Sections \ref{sec:methods} and \ref{sec:theory}, 
respectively,
followed by simulation results in Section \ref{sec:sim} and an application in Section \ref{sec:application}.


\section{High-dimensional vector autoregressive processes with additional structure}
\label{sec:models}

We assume that $\bX(t) = (X_1(t), \dots, X_k(t))_{t=1}^N$ is generated by a $L$-th order vector autoregressive process of the form
\beno
\bX(t) &=& \dsum_{l=1}^L \bA_l\,\bX(t-l) + \bee(t),
\ee
where $\bA_1, \dots, \bA_L$ are $k \times k$ transition matrices and the errors $\bee(t)$ are independent multivariate Gaussian random variables with mean $\bm{0}_k$ and positive-definite variance-covariance matrix $\bSigma$.
We follow \citet{LoWa11} and \citet{BaMi13} and assume that the order $L$ of the vector autoregressive process is either known or can be bounded above and that the vector autoregressive process is stable and thus stationary \citep{Lu07}.
In applications where the order $L$ of the vector autoregressive process is unknown and cannot be bounded above,
cross-validation can be used to select $L$.

\subsection{Additional structure} 

We consider high-dimensional vector autoregressive processes where the number of parameters $p = k^2\, L + k^2$ is much larger than the number of observations $N$.
While high-dimensional vector autoregressive processes are challenging due to the dependent and high-dimensional nature of the data,
in many applications there is additional structure that can be exploited to reduce computing time along with statistical error.
We consider high-dimensional vector autoregressive processes with additional structure in the form of space.
In particular,
we assume that the components $i$ of the vector autoregressive process have positions in the interior of a bounded subset $\mbZ \subset \mR^d$.
The boundedness assumption is motivated by applications:
most spatial structures arising in applications can be represented by bounded subsets of $\mR^d$.
Throughout,
we represent the components of the vector autoregressive process by a mixed graph,
where the nodes represent components,
a directed edge from component $i$ to component $j$ indicates that element $(j, i)$ of at least one of the transition matrices $\bA_1, \dots, \bA_L$ is non-zero,
and an undirected edge between components $i$ and $j$ indicates that elements $(i, j)$ and $(j, i)$ of $\bSigma^{-1}$ are non-zero \citep{Ei12}.
We note that the graphical representation of the model is convenient, 
but not essential: all results reported here could be described in terms of non-zero parameters.

\subsection{Model estimation exploiting additional structure} 
\label{exploit}

If additional structure is available,
such as spatial structure,
model estimation should take advantage of it.

To do so,
observe that the boundedness of $\mZ \subset \mR^d$ implies that there exists $\rho_{\max} < \infty$ such that the Euclidean distance $d(i, j)$ between components $i$ and $j$ satisfies $d(i, j) \leq \rho_{\max}$ for all $(i, j) \in \mN \times \mN$,
where $\mN = \{1, \dots, k\}$ denotes the set of components.
Let $\rho$ be the maximum distance separating two components $(i, j)$ with an edge.
By definition of $\rho$,
for each component $i$,
all edges of $i$ are either in the interior or on the boundary of the closed ball centered at the position of $i$ in $\mZ \subset \mR^d$ with radius $\rho \leq \rho_{\max}$ (see, e.g., Figure \ref{fig:local.dependence}).
In light of the fact that all edges,
i.e., 
all non-zero parameters of all components are within distances $d \leq \rho$,
model estimation of non-zero parameters should be restricted to distances $d \leq \rho$.

\ghoster{
\begin{figure}[t]
\begin{center}
\includegraphics[scale=0.98]{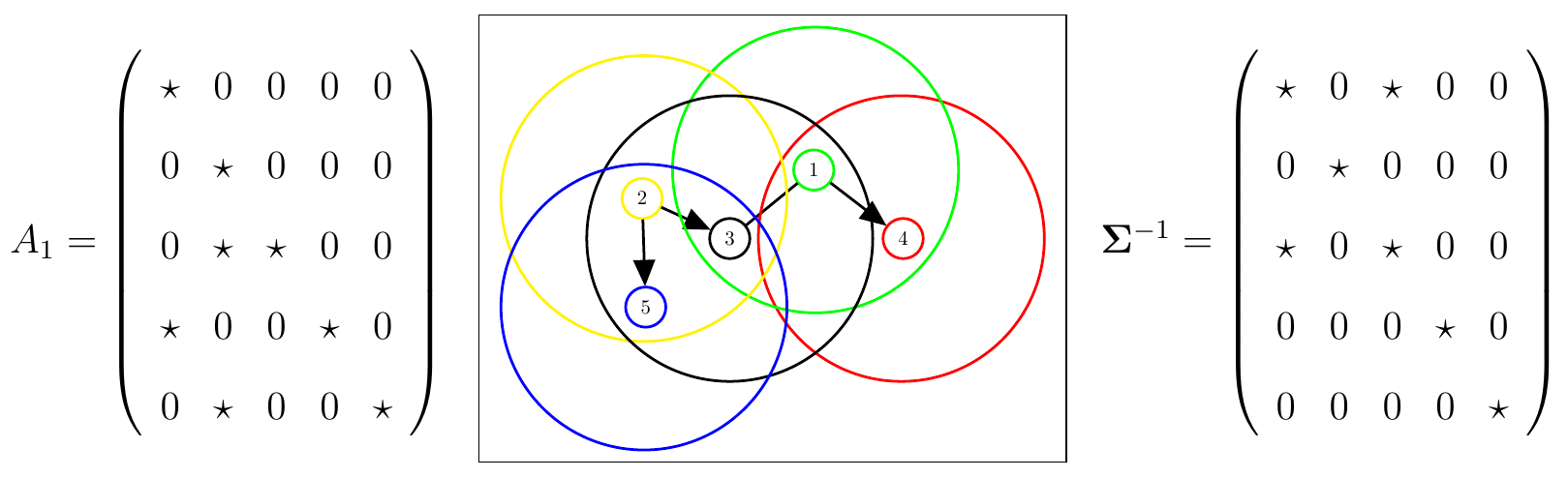}

\caption{
\label{fig:local.dependence}
First-order vector autoregressive process with additional structure: 
nodes represent components of the vector autoregressive process with positions in a bounded subset $\mZ \subset \mR^d$ and edges represent non-zero elements of either $\bA_1$ or $\bSigma^{-1}$. 
The edges of component $i$ are contained in the closed balls with radius $\rho$ centered at the positions of the components.
The elements $\star$ of matrices indicate non-zero elements.}
\end{center}
\end{figure}
}

In practice,
the radius $\rho$ is sometimes known or can be bounded above based on domain knowledge,
but in most cases $\rho$ is unknown and must be estimated.
We introduce methods and theory for estimating $\rho$ in Sections \ref{sec:methods} and \ref{sec:theory} with a view to reducing computing time along with the statistical error of parameter estimators.
It is worth noting that we do not make model assumptions about how the distance between components of the vector autoregressive process affects the dependence between the components: 
all we assume is that components have positions in a bounded subset $\mZ \subset \mR^d$.
Therefore,
the methods can be applied to all vector autoregressive processes with additional structure of the form considered here,
including vector autoregressive processes with $\rho = \rho_{\max}$,
but the greatest reduction in computing time and statistical error is obtained when $\rho \ll \rho_{\max}$ and the components are not too close to each other in $\mZ \subset \mR^d$.

\section{Two-step $\ell_1$-penalized least squares method}
\label{sec:methods}

We introduce a simple two-step $\ell_1$-penalized least squares method that takes advantage of the additional structure considered here.

The two-step $\ell_1$-penalized least squares method is sketched in Table \ref{algorithm}.
It is motivated by the fact that all edges,
i.e., 
all non-zero parameters of all components are within distances $d \leq \rho$,
thus model estimation of non-zero parameters should be restricted to distances $d \leq \rho$.
In practice,
the radius $\rho$ may be unknown.
If the structure of the graph was known,
one could take $\rho$ to be the maximum distance that separates a pair of nodes with an edge.
If the structure of the graph is unknown,
one needs to estimate the graph.
An appealing alternative to estimating the whole graph---which is time-consuming when the set of nodes $\mN$ is large---is to estimate a subgraph by sampling a subset of nodes $\mS$,
estimating the edges of nodes $i \in \mS$,
and then estimating $\rho$ by $\wm$,
defined as the maximum distance that separates a pair of nodes with an estimated edge.
Step 1 estimates the radius $\rho$ by $\wm$ along these lines.
Step 2 estimates the parameters by restricting the estimation of parameters to distances $d \leq \wwm$. 
If the sample in Step 1 is small but well-chosen and the radius $\rho$ is small,
the two-step $\ell_1$-penalized least squares method reduces computing time and statistical error.

\ghoster{
\begin{table}[t]
\begin{center}
\fbox{
\begin{minipage}{5.25in}
\vspace{0.125cm}
\begin{itemize}
\item[1.] If radius $\rho$ is unknown, estimate $\rho$:
\begin{itemize}
\item[1.1] Sample a subset of nodes $\mS$ from the set of nodes $\mN$.
\item[1.2] Estimate edges by regressing nodes $i \in \mS$ on $\{j\, \mid \, j \in \mN \setminus i\}$, i.e., on all other nodes in $\mN$. 
\item[1.3] Estimate radius $\rho$ by $\wwm$, 
the maximum distance that separates a pair of nodes with an estimated edge.
\end{itemize}
\item[2.] Estimate the parameters by using the $\ell_1$-penalized least squares method subject to the constraint that all parameters governing possible edges at distances $d > \wwm$ are $0$.
\end{itemize}
\end{minipage}
}

\s

\caption{
\label{algorithm}
Two-step $\ell_1$-penalized least squares method.}
\end{center}
\end{table}
}

We discuss the implementation of the two-step $\ell_1$-penalized least squares method in Sections \ref{sec:step1} and \ref{sec:step2} and shed light on its theoretical properties in Section \ref{sec:theory}.
Throughout,
we assume that $\bSigma^{-1}$ is diagonal;
extensions to non-diagonal $\bSigma^{-1}$ are possible,
though less attractive on computational grounds \citep[][]{BaMi13}.
We denote by $\norm{.}_1$, $\norm{.}_2$, and $\norm{.}_\infty$ the $\ell_1$, $\ell_2$, and $\ell_\infty$-norm of vectors,
respectively.
The total number of observations is denoted by $M$ and the effective number of observations by $N = M - L + 1$.

\subsection{Step 1}
\label{sec:step1}

If the radius $\rho$ is unknown,
it is estimated in Step 1.

In Step 1.1,
a sample of nodes $\mS$ from the set of nodes $\mN$ is generated by using any sampling design for sampling from finite populations \citep[see, e.g.,][]{Th12}.
Some guidance with respect to sampling designs is provided in Remark \ref{remark7} in Section \ref{sec:theory}.
An example is given in Section \ref{sec:application}. 

In Step 1.2,
edges are estimated by regressing nodes $i \in \mS$ on $\{j\, \mid \, j \in \mN \setminus i\}$ by the $\ell_1$-penalized least squares method of \citet{BaMi13},
which is attractive on both computational and theoretical grounds. 
It is worth noting that regressing sampled nodes $i \in \mS$ on all other sampled nodes in $\mS$ would give rise to omitted variable problems. 
Step 1.2 therefore regresses sampled nodes $i \in \mS$ on {\em all other nodes in $\mN$} rather than {\em all other sampled nodes in $\mS$.} 

To introduce the $\ell_1$-penalized least squares method used in Step 1.2,
note that the conventional $\ell_1$-penalized least squares method estimates the $p=k^2\, L$-dimensional parameter vector $\bbeta_{\mN} = (\bbeta_i)_{i\in\mN}$ corresponding to the vectorized transition matrices $\vv(\bA_1^{\top}, \dots, \bA_L^{\top})$ by
\be 
\label{optimization1a}
\wh\bbeta_{\mN}
&\in& \underset{\bbeta_i,\, i \in \mN}{\argmin} \dsum_{i\in\mN} \left[\dfrac{1}{N}\norm{\bYY_i - \bXX\, \bbeta_i}^2_2 + \lambda_{1}\, \norm{\bbeta_i}_1\right],
\ee
where $\bbeta_i$ denotes the $p_i = k\, L$-dimensional parameter vectors governing possible incoming edges of nodes $i$;
$\bYY_i$ denotes the $i$-th column of the matrix of observations $\bYY = (\bX(M)^{\top}, \dots, \bX(L)^{\top})$;
$\bXX$ denotes the predictors $((\bX(M-1)^{\top}, \dots, \bX(L-1)^{\top})$, $\dots$, $(\bX(M-L)^{\top}, \dots, \bX(0)^{\top}))$;
and $\lambda_{1} > 0$ denotes a regularization parameter.
The $\ell_1$-penalized least squares method used in Step 1.2 applies the same procedure to the subset of nodes $\mS$ and estimates the parameter vector $\bbeta_{\mS} = (\bbeta_i)_{i\in\mS}$ by
\be 
\label{optimization1b}
\wh\bbeta_{\mS}
&\in& \underset{\bbeta_i,\, i \in \mS}{\argmin} \dsum_{i\in\mS} \left[\dfrac{1}{N}\norm{\bYY_i - \bXX\, \bbeta_i}^2_2 + \lambda_{1}\, \norm{\bbeta_i}_1\right].
\ee
The incoming edges of nodes $i \in \mS$ can be inferred from the non-zero pattern of $\wh\bbeta_{\mS} = (\wh\bbeta_i)_{i\in\mS}$.
The radius $\rho$ can be estimated by $\wm$,
the maximum distance that separates a pair of nodes $(j, i) \in \mN\times\mS$ with an estimated edge,
i.e.,
with an estimated non-zero autoregressive coefficient.

\subsection{Step 2}
\label{sec:step2}

In Step 2,
the parameter vector $\bbeta \equiv \bbeta_{\mN}$ is estimated by restricting the $\ell_1$-penalized least squares method to distances $d \leq \wwm$,
i.e.,
the parameter vector $\bbeta$ is estimated by
\be
\label{optimization2}
\wh\bbeta
&\in& \argmin\limits_{\bbeta_i,\, i \in \mN} \dsum_{i\in\mN} \left[\dfrac{1}{N}\norm{\bYY_i - \bXX\, \bbeta_i}^2_2 + \lambda_{2}\, \norm{\bbeta_i}_{1}\right]
\ee
subject to the constraint that all parameters governing possible edges at distances $d > \wm$ are $0$,
where $\lambda_{2} > 0$ is a regularization parameter.

\com An important observation is that the parameter vectors $\bbeta_1, \dots, \bbeta_k$ are variation-independent in the sense that the parameter space of $\bbeta$ is a product space of the form $\mR^{k^2 L} = \mR^{k L} \times \cdots \times \mR^{k\, L}$.
As a result,
optimization problems \eqref{optimization1a}, \eqref{optimization1b}, and \eqref{optimization2} can be decomposed into $k$ separate optimization problems that can be solved in parallel,
thus reducing computing time.

\com The variance-covariance matrix $\bSigma$ can be estimated by using the $\ell_1$-penalized maximum likelihood method of \citet{BaMi13}.
However,
the $\ell_1$-penalized maximum likelihood method is more expensive in terms of computing time than the $\ell_1$-penalized least squares method.

\section{Theoretical properties}
\label{sec:theory}
	
We provide non-asymptotic bounds on the statistical error of parameter estimators in high-\linebreak
dimensional settings and show that the two-step $\ell_1$-penalized least squares method reduces the statistical error.
To facilitate the discussion,
we follow \citet{LoWa11} and \citet{BaMi13} by expressing optimization problems \eqref{optimization1a}, \eqref{optimization1b}, and \eqref{optimization2} as $M$-estimation problems of the form
\beno
\widehat{\bbeta}
&\in& \argmin\limits_{\bbeta\, \in\, \mC} \left[-2\, \bbeta^\top\widehat{\bgamma}+\bbeta^\top\, \widehat{\bGamma}\, \bbeta + \lambda\, \norm{\bbeta}_1\right],
\ee
where $\mC$ is a subset of $\mbR^p$ that depends on the constraints imposed by optimization problems \eqref{optimization1a}, \eqref{optimization1b}, and \eqref{optimization2}, 
$\widehat{\bgamma} = (\bI \otimes  \bXX^\top)\, \vv(\bYY) / N$,
and $\widehat{\bGamma} = (\bI \otimes \bXX^\top\bXX) / N$,
where $\bI$ denotes the identity matrix of suitable order and $\otimes$ denotes the Kronecker product.

{\em Notation.}
Throughout,
we assume that the elements of $\bbeta$ and $\bgamma$ are ordered according to distance and denote by $\bbeta_{[d_1, d_2]}$ and $\bgamma_{[d_1, d_2]}$ the subvectors of $\bbeta$ and $\bgamma$ corresponding to parameters governing possible edges at distances $d \in [d_1, d_2]$,
respectively,
where $0 \leq d_1 \leq d_2$.
The rows and columns of $\bGamma$ are ordered in accordance.
Denote by $p(0, d_2)$ the total number of parameters governing possible edges at distances $d \in [0, d_2]$ and by $p(d_1, d_2)$ the total number of parameters governing possible edges at distances $d \in (d_1, d_2]$,
where $0 < d_1 \leq d_2$.
Let $\wh\bbeta$ be the estimator of the true parameter vector $\bbeta^\star$ obtained by the two-step $\ell_1$-penalized least squares method.
Denote by $\mbS$ the support of $\bbeta^\star$ and by $s$ the size of support $\mbS$.
Let $\delta > 0$ and $\mS(\delta)$ be the subset of nodes with incoming edges at distances $d \in [\rho-\delta, \rho]$. 
We denote by $c_0, c_1, c_2 > 0$ unspecified constants.

We assume that the following conditions hold. 
The first assumption is a restricted eigenvalue condition,
whereas the second condition is a deviation condition.
Both conditions are conventional and hold with high probability \citep[][]{LoWa11,BaMi13}.

\assumption
\label{ass1}
$\widehat{\bGamma}$ satisfies the restricted eigenvalue condition with curvature $\alpha > 0$ and tolerance $\tau > 0$ provided $s\, \tau \leq \alpha / 32$ and 
\beno
\bm{b}^\top\, \widehat{\bGamma}\, \bm{b} &\geq& \alpha\, \norm{\bm{b}}_2^2 - \tau\, \norm{\bm{b}}^2_1 &\mbox{for all}& \bm{b} \in \mathbb{R}^p.
\ee

\assumption
\label{ass2}
There exists a deterministic function $\mQ > 0$ such that $\wh\bgamma$ and $\wh\bGamma$ satisfy
\beno
\norm{\widehat{\bgamma} - \widehat{\bGamma}\, \bbeta^\star}_{\infty} &\leq& \mQ\, \sqrt{\dfrac{\log p}{N}}.
\ee

\hide{
 and specified the form of functions $\alpha$, $\tau$, and $\mQ$.
Here, 
the form of these functions is immaterial and the existence and scalings of these functions is all we need.
}
\s

The following theorems show that the two-step $\ell_1$-penalized least squares method reduces the statistical error of parameter estimators without making model assumptions about how the distance between the components of the vector autoregressive process affects the dependence between the components.
We start with the case where $\rho$ is either known or can be bounded above based on domain knowledge (Theorem \ref{theorem.known}) and then turn to the case of unknown $\rho$ (Theorem \ref{theorem.unknown}).
To streamline the presentation,
Theorem \ref{theorem.known} focuses on known $\rho$,
but the extension to bounded $\rho$ is straightforward.

{\em 
\theorem
\label{theorem.known}
Consider $N \geq c_0\, s \log p$ ($c_0 > 1$) observations from a stable $L$-th order vector autoregressive process with radius $\rho > 0$.
Suppose that $\rho$ is known and that the regularization parameter $\lambda_2$ in the second step of the two-step $\ell_1$-penalized least squares method satisfies
\be
\label{lambda0}
\lambda_2
&\geq& 4\, \mQ\, \sqrt{\dfrac{\log p(0,\rho)}{N}}.
\ee
Then,
with at least probability
\be
\label{prob.b}
1 - 2\, \exp(- c_1\, N) - 6\, \exp(- c_2 \log p(0,\rho)),
\ee
the $\ell_2$-error of estimator $\wh\bbeta$ of $\bbeta^\star$ is bounded above by
\beno
\norm{\wh\bbeta - \bbeta^\star}_2
&\leq& \dfrac{16\, \sqrt{s}\, \lambda_{2}}{\alpha}.
\ee
}

We compare the statistical error and computing time of the two-step $\ell_1$-penalized least squares method to existing high-dimensional methods.

\com
\label{com.1}
{\em Comparison in terms of statistical error.}
Among the existing approaches,
the most attractive approach is the $\ell_1$-penalized least squares method of \citet{BaMi13},
because it has computational advantages and its theoretical properties are well-understood.
Suppose that $\bbeta^\star$ is estimated by the two-step $\ell_1$-penalized least squares method with known $\rho$ with $\lambda_2 = 4\, \mQ\, \sqrt{\log p(0,\rho) / N}$.
Then,
with high probability,
\begin{equation}
\begin{array}{ccccccc}
\nonumber
\norm{\wh\bbeta - \bbeta^\star}_2
&\leq& \underbrace{\dfrac{64}{\alpha} \mQ \sqrt{\dfrac{s \log p(0,\rho)}{N}}}
&\leq& \underbrace{\dfrac{64}{\alpha} \mQ \sqrt{\dfrac{s \log p}{N}}},\s
\\
&& \mbox{two-step $\ell_1$-least squares} && \mbox{$\ell_1$-least squares}
\end{array}
\end{equation}
because $p(0,\rho) = \sum_{i=1}^k n_i(\rho)\, L \leq p = k^2\, L$,
where $p(0,\rho)$ is the total number of parameters governing possible edges at distances $d \in [0, \rho]$ and $n_i(\rho)$ is the number of components $j \in \mN \setminus i$ within distance $d(i, j) \leq \rho$ of component $i$.
The error bounds show that restricting model estimation to distances $d \leq \rho$ reduces the $\ell_2$-error of $\wh\bbeta$.

\com 
\label{com.compu}
{\em Comparison in terms of computing time.}
In terms of computing time,
the two-step $\ell_1$-penalized least squares method with known (bounded) $\rho$ tends to be superior to the $\ell_1$-penalized least squares method:
while the $\ell_1$-penalized least squares method amounts to running $k$ regressions with $k\, L$ predictors,
the two-step $\ell_1$-penalized least squares method with known (bounded) $\rho$ amounts to running $k$ regressions with $\max_{1 \leq i \leq k} n_i(\rho)\, L$ predictors,
where $n_i(\rho)$ is the number of components $j \in \mN \setminus i$ within distance $d(i, j) \leq \rho$ of component $i$.
If $\max_{1 \leq i \leq k} n_i(\rho)\, \ll k$,
the two-step $\ell_1$-penalized least squares method with known (bounded) $\rho$ is many times faster than the $\ell_1$-penalized least squares method and can thus be applied to much larger data sets.

\s

We turn to the case where $\rho$ is unknown.
Choose $\delta > 0$ as small as desired and consider the estimator $\wh\bbeta_{[0, \rho-\delta]}$ of the parameter vector $\bbeta_{[0, \rho-\delta]}^\star$ governing possible edges to nodes in the interior of the balls centered at the positions of nodes,
which---in most applications---are the parameters of primary interest.
Theorem \ref{theorem.unknown} bounds the statistical error of the estimator $\wh\bbeta_{[0, \rho-\delta]}$ of the parameter vector $\bbeta_{[0, \rho-\delta]}^\star$ for all $\delta > 0$.

{\em
\begin{theorem}
\label{theorem.unknown}
Consider $N \geq c_0\, s \log p$ ($c_0 > 1$) observations from a stable $L$-th order vector autoregressive process with radius $\rho > 0$.
Assume that components $i$ are sampled independently with probabilities $0 < \theta_i < 1$ and that the minimum signal strength is $\beta_{\min}^\star = \min_{i \in \supp} |\beta_i^\star| \geq 32\, \sqrt{s}\, \lambda_1 / \alpha > 0$.
Choose any $\delta > 0$,
however small,
and assume that the regularization parameters $\lambda_{1}$ and $\lambda_{2}$ in the first and second step of the two-step $\ell_1$-penalized least squares method satisfy
\be
\label{lambda1}
\lambda_{1}
&\geq& 4\, \mQ\, \sqrt{\dfrac{\log p}{N}}
\ee
and
\be
\label{lambda2}
\lambda_{2}
&\geq& 4\, \mQ\, \sqrt{\dfrac{\log p(0,\rho-\delta)}{N}},
\ee
respectively.
Then,
for all $\delta > 0$,
with at least probability
\be
\label{prob.b}
1 - 4 \exp(- c_1 N) - 12 \exp(- c_2 \log p(0,\rho-\delta)) - \exp\left(-\dsum_{i\in\mS(\delta)} \theta_i\right),
\ee
the $\ell_2$-error of the estimator $\wh\bbeta_{[0,\rho-\delta]}$ of the parameter vector $\bbeta_{[0,\rho-\delta]}^\star$ is bounded above by
\beno
\norm{\wh\bbeta_{[0, \rho-\delta]} - \bbeta^\star_{[0, \rho-\delta]}}_2
&\leq& \dfrac{16\, \sqrt{s}\, \lambda_{2}}{\alpha}.
\ee
\end{theorem}
}

\com {\em Statistical error.}
\label{com.theorem2}
The so-called beta-min condition in Theorem \ref{theorem.unknown},
which asserts that the non-zero elements of $\bbeta^\star$ cannot be too small,
is common in the literature on high-dimensional variable selection and graphical models \citep[see, e.g.,][Section 7.4]{BuGe11}.
It is needed to make sure that the 
edges of sampled nodes can be recovered with high probability,
which in turn is needed to estimate the radius $\rho$.
Theorem \ref{theorem.unknown} shows that the statistical error of estimators of the parameter vector $\bbeta_{[0, \rho-\delta]}^\star$ governing possible edges in the interior of the balls---which,
in most applications,
are the parameters of primary interest---is small when the number of observations $N$ is large relative to the size of the support $s$ and the number of parameters $p(0,\rho-\delta)$.
The statistical error of the estimator $\wh\bbeta$ of the whole parameter vector $\bbeta$ is more complicated.
On the one hand,
if $\rho$ is overestimated in Step 1,
the error bound of the estimator $\wh\bbeta$ in Step 2 is at most as large as the error bound of the estimator $\wh\bbeta$ under the $\ell_1$-penalized least squares method,
which follows from Theorem \ref{theorem.known} and Remark \ref{com.1}.
On the other hand,
if $\rho$ is underestimated in Step 1,
the parameter vector $\bbeta_{(\rho-\delta,\rho]}^\star$ governing possible edges close to the boundary of the balls centered at the positions of nodes is not estimated in Step 2,
thus the error bound of the estimator $\wh\bbeta$ in Step 2 depends on the $\ell_2$-norm of $\bbeta_{(\rho-\delta,\rho]}^\star$.
\hide{
which follows from the triangle inequality:
\beno
\label{mybound2}
\norm{\wh\bbeta - \bbeta^\star}_2
\hide{
&\leq& \norm{\wh\bbeta - \bbeta^\star_{0}}_2 + \norm{\bbeta^\star_{0} - \bbeta^\star}_2\s
\\
}
&\leq& 
\hide{
\norm{\wh\bbeta_{[0,\rho-\delta]} - \bbeta^\star_{[0,\rho-\delta]}}_2 + \norm{\wh\bbeta_{(\rho-\delta, \rho]} - \bbeta^\star_{(\rho-\delta, \rho]}}_2
&=& 
}
\norm{\wh\bbeta_{[0,\rho-\delta]} - \bbeta^\star_{[0,\rho-\delta]}}_2 + \norm{\bbeta^\star_{(\rho-\delta, \rho]}}_2.
\ee
}

\hide{
e.g.,
suppose that $\rho$ is estimated by $\wm = \rho - \delta$ in the first step,
where $\delta > 0$;
then the triangle inequality shows that the $\ell_2$-error of the estimator $\wh\bbeta$ of $\bbeta^\star$ can be decomposed into the $\ell_2$-error of estimating $\bbeta^\star_{[0,\rho-\delta]}$ and the $\ell_2$-error of estimating $\bbeta^\star_{(\rho-\delta, \rho]}$:
\beno
\label{mybound2}
\norm{\wh\bbeta - \bbeta^\star}_2
\hide{
&\leq& \norm{\wh\bbeta - \bbeta^\star_{0}}_2 + \norm{\bbeta^\star_{0} - \bbeta^\star}_2\s
\\
}
&\leq& 
\hide{
\norm{\wh\bbeta_{[0,\rho-\delta]} - \bbeta^\star_{[0,\rho-\delta]}}_2 + \norm{\wh\bbeta_{(\rho-\delta, \rho]} - \bbeta^\star_{(\rho-\delta, \rho]}}_2
&=& 
}
\norm{\wh\bbeta_{[0,\rho-\delta]} - \bbeta^\star_{[0,\rho-\delta]}}_2 + \norm{\bbeta^\star_{(\rho-\delta, \rho]}}_2,
\hide{
\\
&\leq& \dfrac{16\, \sqrt{s}\, \lambda_{2}}{\alpha} + \norm{\bbeta^\star_{(\rho-\delta, \rho]}}_2.
}
\ee
where $\norm{\bbeta^\star_{(\rho-\delta, \rho]}}_2$ denotes the $\ell_2$-norm of the parameter subvector governing possible edges at distances $d \in (\rho-\delta, \rho]$,
i.e., 
possible edges from nodes close to the boundary of the balls centered at the positions of nodes.
}
\hide{
If the $\ell_2$-norm of parameter vector $\bbeta^\star_{(\rho-\delta, \rho]}$ is small,
as one would expect when dependence is more short- than long-range,
the bounds on the error of $\wh\bbeta$ may be small even when $\rho$ is underestimated.
}
\hide{
However, in any case, the error bounds of the estimator $\wh\bbeta_{[0,\rho-\delta]}$ of the parameters governing edges in the interior of the the balls are small with high probability and decay as a function of the number of observations $N$.
}

\com {\em Computing time.}
In terms of computing time,
the two-step $\ell_1$-penalized least squares method amounts to running $|\mS|$ regressions with $k\, L$ predictors in Step 1 and $k$ regressions with $\max_{1 \leq i \leq k} n_i(\wm)\, L$ predictors in Step 2 of the two-step $\ell_1$-penalized least squares method,
where $\wm$ is the estimate of $\rho$ obtained in Step 1.
Therefore,
as long as the sample is small but well-chosen and the radius is short,
the two-step $\ell_1$-penalized least squares method outperforms $\ell_1$-penalized least squares method.


\hide{
According to \cite{MeBu10} there exists a constant $C$ such that the algorithmic complexity of LASSO with stability selection for a regression problem with $k\,L$ predictors and N observations  is $C\, N\, k\, L \, \min(N, k\, L)$. Since the conventional approach  is simply $k$ separate full LASSO programs its complexity is $C \,N \,k^3\,L^2$ if we assume $N > k \, L$ \alert{this is a better case for us since we get more reduction in complexity if $N > k \, L$} . Whereas for the local approach, we sample $k^{f_1}$ nodes in step 1 (obviously, $f_1 < 1$), therefore the complexity of first step is $C N \,k^{1+2f_1}\,L^2$. Let $f_2$ be such that on average we have $k^{f_2} \, L$ predictors left within the radius in Step 2 (again $f_2 < 1$). Then the complexity of the second step is  $C N \,k^{1+2f_2}\,L^2$ and the resulting complexity of local approach is $C N \,L^2 \,(k^{1+2f_1}+k^{1+2f_2})$. In practice, we had $f_1 \approx 0.4 - 0.6$ and $ f_2\approx 0.7 - 0.8$ throughout simulation study and application.
}

\com {\em Sampling.}
\label{remark7}
Theorem \ref{theorem.unknown} shows that,
for any given $\delta > 0$,
the probability of the event that $\norm{\wh\bbeta_{[0, \rho-\delta]} - \bbeta^\star_{[0, \rho-\delta]}}_2$ is small depends on the term $\exp(-\sum_{i\in\mS(\delta)} \theta_i)$:
that is,
it depends on 
(a) the size of $\mS(\delta)$ and (b) the sample inclusion probabilities $\theta_i$ of nodes $i \in \mS(\delta)$,
i.e.,
nodes with incoming edges at distances $d \in [\rho-\delta, \rho]$.
The first factor is outside of the control of investigators,
whereas the second factor is under the control of investigators.
The fact that the probability of the event of interest depends on the sample inclusion probabilities $\theta_i$ of nodes $i \in \mS(\delta)$ rather than nodes $i \in \mN \setminus \mS(\delta)$ shows that one needs to sample nodes $i \in \mS(\delta)$ rather than nodes $i \in \mN \setminus \mS(\delta)$ with high probability.
In other words,
non-uniform sampling designs that sample nodes with long-distance edges with high probability are preferable to uniform sampling designs and the number of sampled nodes with long-distance edges is more important than the total number of sampled nodes.
Therefore,
if prior knowledge is available about which nodes may have long-distance edges,
it should be incorporated into the sampling design.
Such prior knowledge is available in a number of spatio-temporal applications:
e.g.,
in studies of air pollution,
it is well-known that industrial and metropolitan areas tend to spread air pollution to surrounding areas and that some geographical conditions in combination with wind conditions facilitate long-distance transport of pollutants.
Thus,
pollution monitors in industrial and metropolitan areas and other areas suspected of facilitating long-distance transport of pollutants should be sampled with high probability.
In the application in Section 6,
we sample pollution monitors in the 15 most polluted cities in the U.S.\ with high probability and others with low probability.

\section{Simulation results}
\label{sec:sim}

We compare the two-step $\ell_1$-penalized least squares method with known $\rho$ and unknown $\rho$ to the $\ell_1$-penalized least squares method of \citet{BaMi13},
which is the most attractive high-dimensional method available,
as discussed in Remark \ref{com.1} in Section \ref{sec:theory}.
We compare the methods in terms of statistical error and computing time.
Throughout,
we use stability selection \citep{MeBu10} to sidestep the problem that the choice of the regularization parameters $\lambda_1$ and $\lambda_2$ in the first and second step of the two-step $\ell_1$-penalized estimation method depends on the unknown values of $\bbeta^\star$ and $\bSigma$.
We followed the guidelines of \citet{MeBu10} concerning the choice of tuning parameters of stability selection.
The {\tt R} source code we used is contained in the supplementary archive.

\ghoster{
\begin{table}[t]
\begin{center}
\begin{tabular}{|c|c|c|c|c|}
\hline
\multicolumn{1}{|c}{}& & \multicolumn{1}{|c|}{$k = 100$} & \multicolumn{1}{|c|}{$k = 200$} &\multicolumn{1}{|c|}{$k = 300$}\\
\cline{1-5}
\multirow{3}{*}{AUROC} & Least squares & $.994\; (.005)$ & $.968\; (.013)$ & $.867\; (.033)$\\
\cline{2-5}
& Two-step least squares & $.987\; (.016)$ & $.988\; (.011)$ & $.960\; (.021)$\\
\cline{2-5}
& Oracle two-step least squares & $.999\; (.001)$ & $.996\; (.003)$ & $.969\; (.019)$\\
\hline
\multirow{3}{*}{Estimation error} & Least squares & $.374\; (.026)$ & $.525\; (.032)$ & $.714\; (.043)$\\
\cline{2-5}
& Two-step least squares & $.343\; (.028)$ & $.492\; (.037) $& $.666\; (.052)$\\
\cline{2-5}
& Oracle two-step least squares & $.324\; (.019)$ & $.479\; (.032)$ &  $.655\; (.052)$\\
\hline
\multirow{3}{*}{Fraction of FP} & Least squares & $.003\; (.000)$ & $.003\; (.001)$ & $.005\; (.000)$\\
\cline{2-5}
& Two-step least squares & $.001\; (.000)$ & $.002\; (.000)$ & $.004\; (.001)$\\
\cline{2-5}
& Oracle two-step least squares & $.001\; (.000)$ & $.002\; (.000)$ & $.004\; (.001)$\\
\hline
\multirow{3}{*}{Fraction of FN} & Least squares & $.054\; (.016)$ & $.105\; (.028)$ & $.291\; (.058)$\\
\cline{2-5}
& Two-step least squares & $.034\; (.022)$ & $.052\; (.036)$ & $.156\; (.068)$\\
\cline{2-5}
& Oracle two-step least squares & $.018\; (.013)$ & $.033\; (.018)$ & $.133\; (.062)$\\
\cline{1-5}
\end{tabular}

\s

\caption{
\label{table:simulations}
Comparison of the $\ell_1$-penalized least squares method,
the two-step $\ell_1$-penalized least squares method with unknown $\rho$,
and the oracle two-step $\ell_1$-penalized least squares method with known $\rho$.
Monte Carlo standard deviations are given in parentheses.}
\end{center}
\end{table}
}
To shed light on the statistical error of the methods,
we consider three high-dimensional scenarios with $N = 150$ ($k = 100$), $N = 300$ ($k = 200$), and $N = 450$ ($k = 300$) observations;
note that $p = k^2\, L \gg N$ in all three cases.
For each scenario,
we generated data from a VAR$(1)$ process with $k \times k$ transition matrix $\bA \equiv \bA_1$ with $2\%$ sparsity and overlapping neighborhoods.
The overlapping neighborhoods are generated as follows:
we sample 5 ($k = 100$), 10 ($k = 200$), and 15 ($k = 300$) points from the Uniform distribution on a two-dimensional square.
The sampled points are considered to be centers of neighborhoods and,
for each neighborhood,
we sample 20 points from a bivariate Gaussian centered at the neighborhood center.
Then edges are generated so that 90\% of all edges are within neighborhoods and 10\% are between neighborhoods,
subject to the constraint that between-neighborhood edges are at distances less than the 30\% quantile of the empirical distribution of distances.
We compare the methods
in terms of
(a) model selection error: 
the area under the receiving operator characteristic curve (AUROC);
the fraction of false-positive (FP) and false-negative (FN) edges;
and (b) model estimation error: the relative estimation accuracy measured by $||\bA-\widehat{\bA}||_F/||\bA||_F$,
where $||\bA||_F = \sqrt{\mbox{tr}(\bA^\top \bA)}$.
In Table \ref{table:simulations},
we report the results based on 1,000 Monte Carlo simulations along with Monte Carlo standard deviations. 
It is not surprising that the oracle two-step $\ell_1$-penalized least squares method with known $\rho$ seems to perform best,
but the two-step $\ell_1$-penalized least squares method with unknown $\rho$ seems to be close.
Both seem to outperform the $\ell_1$-penalized least squares method.
In Figure \ref{fig:AUROC},
we assess the impact of the number of observations $N$ on model selection error in terms of AUROC using $k = 200$ components.
It is evident that the two-step $\ell_1$-penalized least squares method with unknown $\rho$ outperforms the $\ell_1$-penalized least squares method even when $N$ is as small as $100$.


\ghoster{
\begin{figure}[t]
\begin{center}
\includegraphics[width=0.6\textwidth]{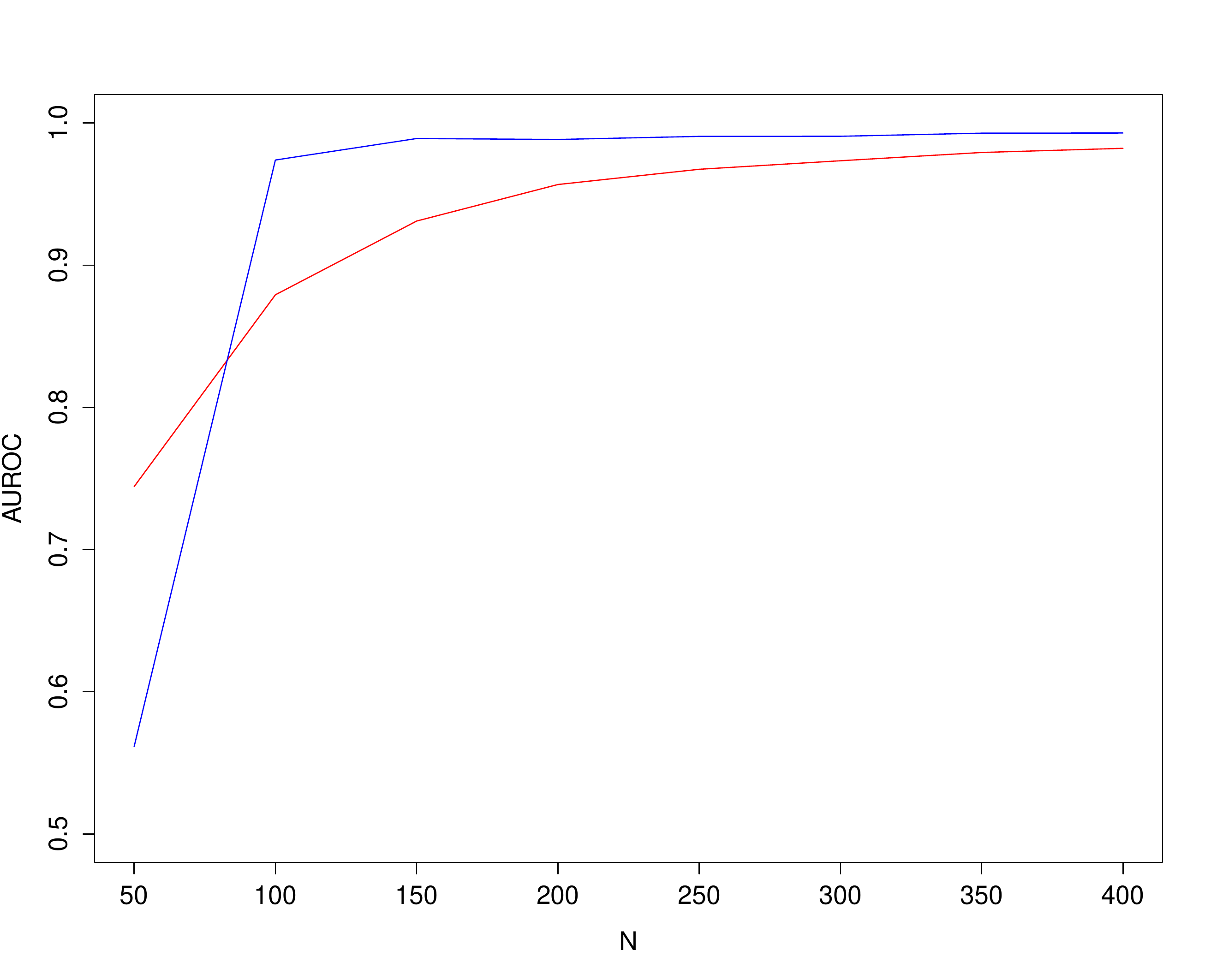}
\caption{
\label{fig:AUROC}
AUROC plotted against number of observations $N$ using $k = 200$ components.
The blue and red line correspond to the two-step $\ell_1$-penalized least squares method with unknown $\rho$ and the $\ell_1$-penalized least squares method, respectively.}
\end{center}
\end{figure}
}

\ghoster{
\begin{figure}[t]
\centering
\includegraphics[width=1\textwidth]{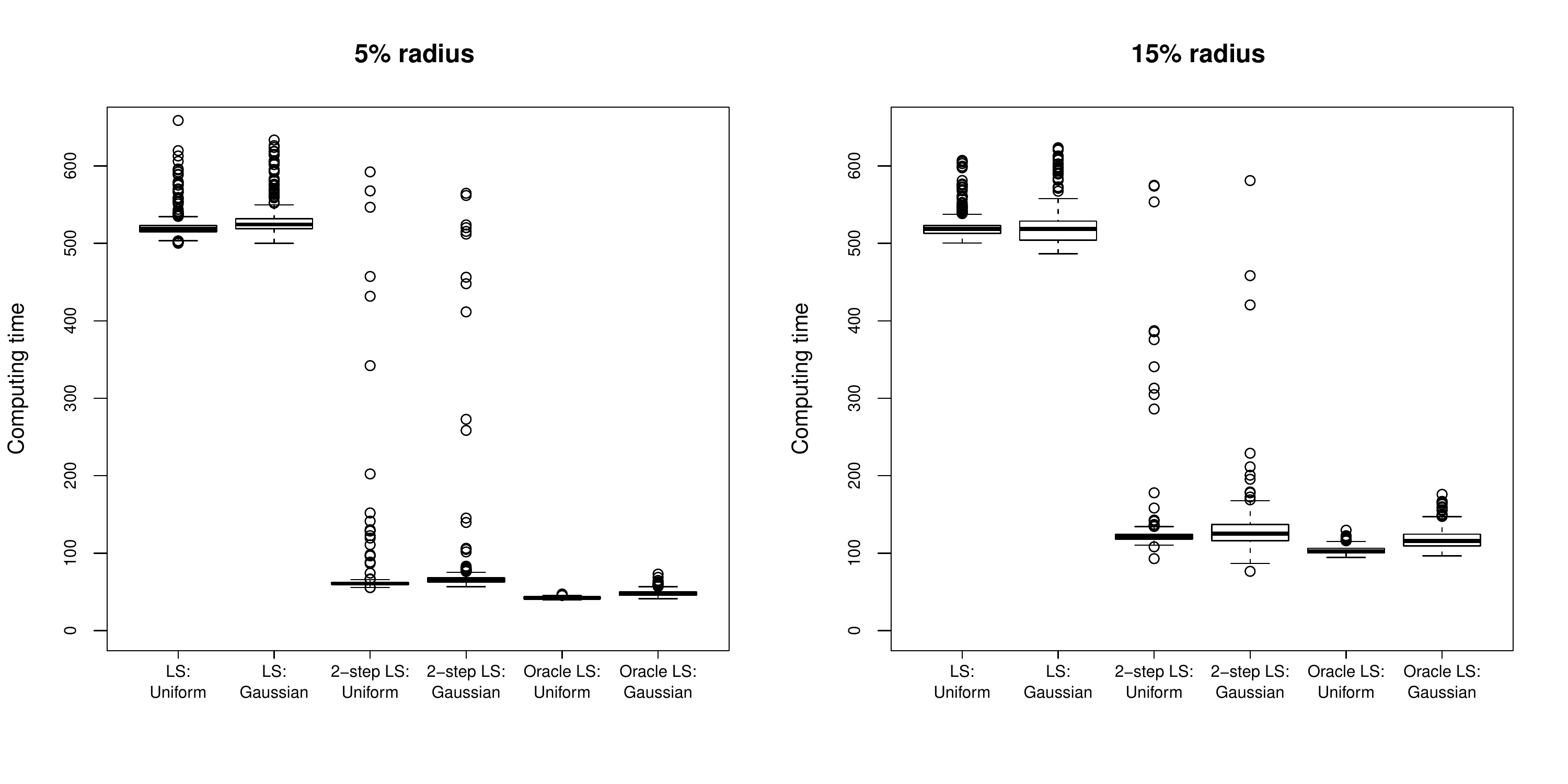}
\caption{Computing time in seconds of the $\ell_1$-penalized least squares method (LS),
the two-step $\ell_1$-penalized least squares method with unknown $\rho$ ($2$-step LS),
and the oracle two-step $\ell_1$-penalized least squares method with known $\rho$ (Oracle LS)
in two spatial settings (Uniform and Gaussian) with small and moderate radius (5\% and 15\%).
\label{boxplots1}
}
\end{figure}
}
To compare the methods in terms of computing time,
we consider $k = 400$ time series governed by a VAR$(1)$ process with a $400 \times 400$ transition matrix $\bA \equiv \bA_1$ with $1\%$ sparsity in the high-dimensional setting where $p = $ 160,000 $\gg N = 600$.
To assess the impact of the spatial structure and the radius on computing time,
we compare the methods in two spatial settings and, for each spatial setting, we use a small and a moderate radius $\rho$.
The two spatial settings are generated by two processes.
The first generating process,
called Uniform generating process, 
generates spatial positions of time series by sampling 400 points from the Uniform distribution on a two-dimensional square.
The second generating process,
called Gaussian generating process, 
generates spatial positions of time series by first sampling 20 points from the Uniform distribution on a two-dimensional square.
The 20 points are used as centers of 20 bivariate Gaussians and from each bivariate Gaussian 20 points are sampled.
For each spatial structure,
we select a small and a moderate radius.
To make sure that the balls centered at the locations of the time series contain a non-negligible fraction of possible edges,
we use the 5\% and 15\% quantile of the empirical distribution of the distances as small and moderate radius,
respectively.
The corresponding radii are called ``5\% radius" and ``15\% radius",
but note that the resulting radius varies from data set to data set,
depending on the spatial positions of the time series.
Conditional on the locations of the $k = 400$ time series, 
we generate $N = 600$ observations from a VAR$(1)$ process with $400 \times 400$ transition matrix $\bA \equiv \bA_1$ with $1\%$ sparsity.
The results based on 500 Monte Carlo simulations are presented in Figure \ref{boxplots1}.
The figure demonstrates that the two-step $\ell_1$-penalized least squares method with known $\rho$ and unknown $\rho$ outperforms the $\ell_1$-penalized least squares method in terms of computing time by a factor of close to 10 (5\% radius) and 5 (15\% radius).
The two-step $\ell_1$-penalized least squares with unknown $\rho$ is almost as fast as the oracle version with known $\rho$,
demonstrating that estimating $\rho$ rather than knowing $\rho$ comes at a cost,
but the cost seems to be low,
with the exception of the rare cases where $\rho$ is overestimated by a non-negligible amount.
The impact of the spatial structure on the computing time seems to be small,
but increasing the radius increases the computing time visibly.

\section{Application to air pollution in the U.S.A.}
\label{sec:application}

\ghoster{
\begin{figure}[t]
\centering
\includegraphics[width=0.4\textwidth]{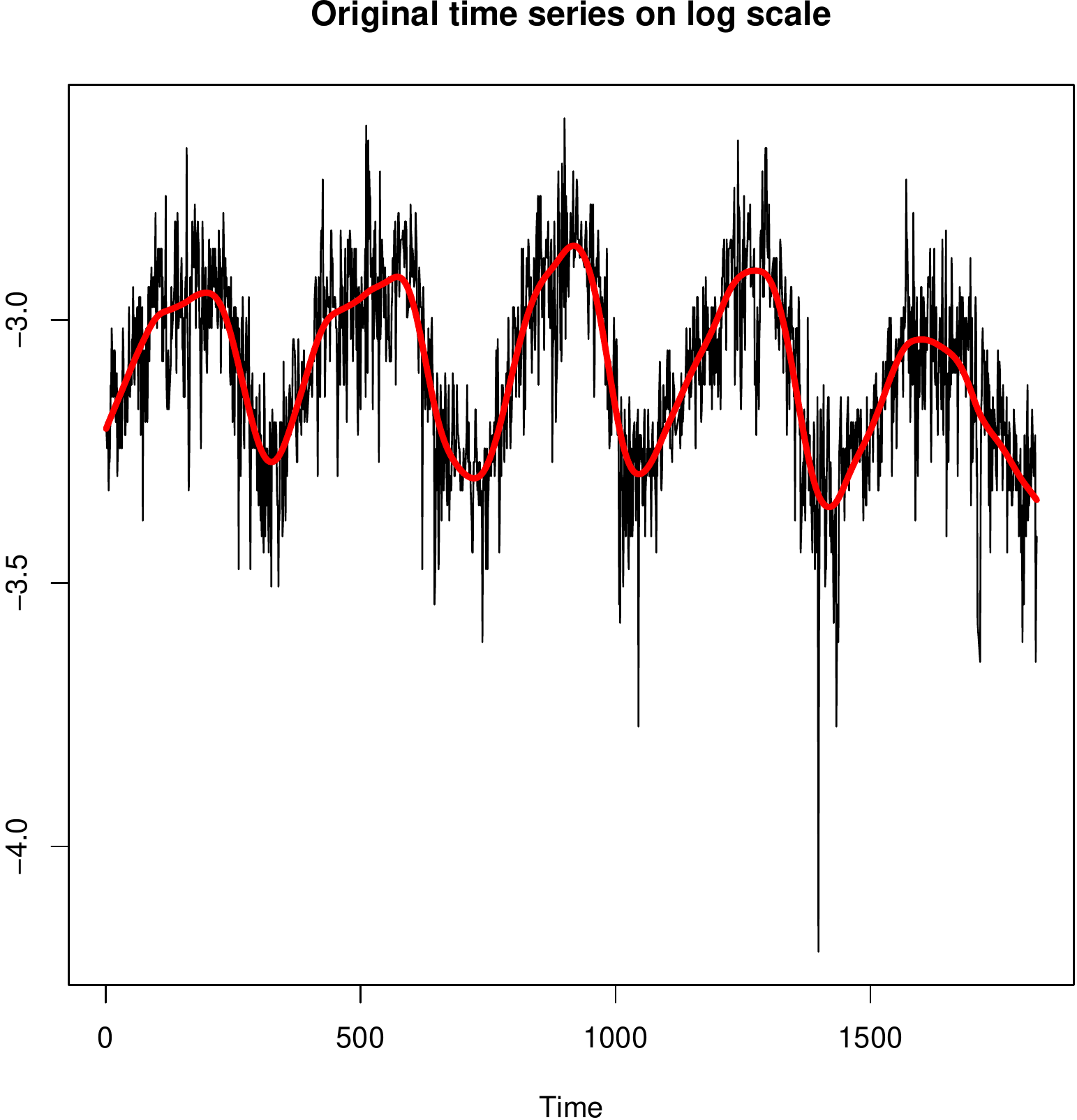}\hspace{10mm}
\includegraphics[width=0.4\textwidth]{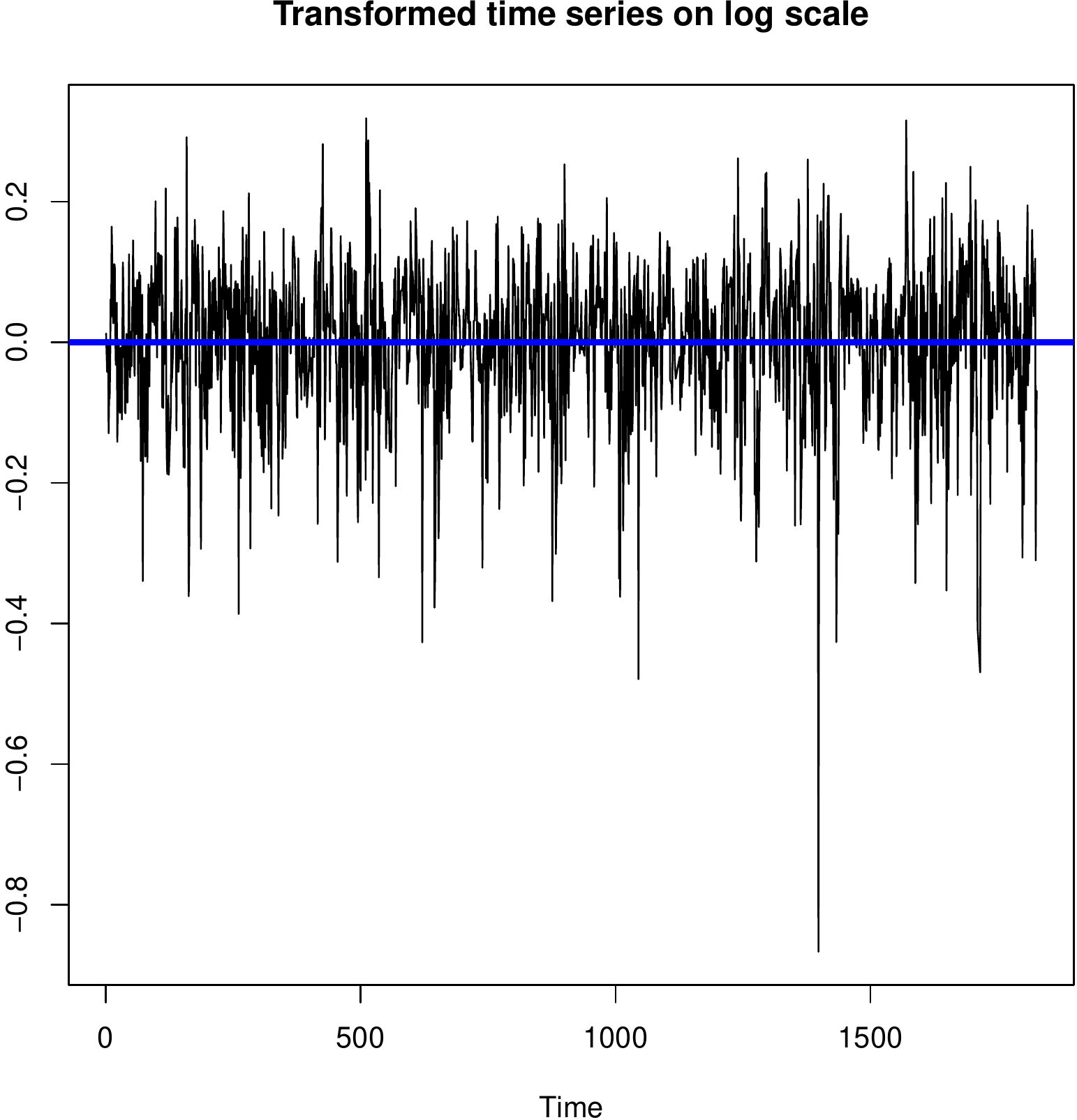}
\caption{Example of Ozone time series consisting of $N =$ 1,826 observations of Ozone levels between January 2010 and December 2014 in its original form and transformed form,
both on the log scale.
The figure on the left-hand side shows the original log Ozone time series.
The $5$ summers increase the log Ozone levels while the $5$ winters decrease them.
The red curve is the fitted cubic spline that captures the seasonal ups and downs.
The figure on the right-hand side shows the transformed log Ozone time series.
The blue line is the mean of the $N =$ 1,826 observations.
}
\label{fig:spline}
\end{figure}
}

Air pollution is an important health concern.
The \citet{ALA15} states that in the U.S.A.\ alone almost 138.5 million people live in areas where air pollution makes breathing dangerous.
Air pollution has been associated with cardiac arrest \citep{EnRa13},
lung disease \citep{Ho13},
and cancer \citep{ChWa15},
and the \citet{WHO14} attributed more than $7$ million deaths in 2012 alone to air pollution.

We exploit the two-step $\ell_1$-penalized least squares method to contribute to the understanding of the 24-hour transport of air pollution across space by using data from the U.S.\ Environmental Protection Agency obtained from
\begin{center}
\url{http://www.epa.gov/airdata/ad_data_daily.html}.
\end{center}
The data are contained in the supplementary archive along with all {\tt R} source code we used to analyze the data. 
Throughout the section,
we use VAR$(1)$ processes,
because Ozone and other pollutants tend to decompose fast.
An additional advantage of using VAR$(1)$ processes is that we have ground truth on the 24--72 hour transport of Ozone in the sense that we have an upper bound on the spatial distance Ozone is known to travel in 24--72 hours \citep[see, e.g.,][]{Raetal97}.
We first take a bird's eye view at air pollution in the U.S.A.\ (Section \ref{app1}) and then zoom in on the Gulf of Mexico region (Section \ref{app2}),
one of the most monitored regions in the U.S.A.

\subsection{A bird's eye view: air pollution in the U.S.A.}
\label{app1}

We consider daily measurements of 8-hour maximum concentration of Ozone ($O_3$) recorded by monitors across the U.S.A.
The data set consists of $N =$ 1,826 observations of Ozone levels recorded by $k =$ 444 monitors between January 2010 and December 2014.
All monitors contain less than 10\% of missing values and we impute the missing values by univariate linear interpolation.
Ozone concentrations were log-transformed and a cubic spline was fitted to each Ozone time series to capture the seasonal ups and downs.
We subtract the fitted cubic splines from the log-transformed Ozone time series and use the residual time series as data.
An example of a Ozone time series in its original and transformed form is shown in Figure \ref{fig:spline}.

We estimate the model by using the two-step $\ell_1$-penalized least squares method, 
using stability selection \citep{MeBu10} to sidestep the problem that the choice of the regularization parameters $\lambda_1$ and $\lambda_2$ in the first and second step of the two-step $\ell_1$-penalized estimation method depends on the unknown values of $\bbeta^\star$ and $\bSigma$.
In Step 1, 
we include pollution monitors in the $15$ most polluted cities in the U.S.A.\ in 2015---according to the website of the American Lung Association---with probability $.99$ and other pollution monitors with probability $.01$.
We excluded 91 pollution monitors in sparsely monitored regions and regions with known omitted monitors---omitted due to a large fraction of missing data---from the sample out of the concern that such monitors may give rise to spurious edges. 
Most of those monitors are located in sparsely populated and mountainous regions in the Midwest and West.

\ghoster{
\begin{figure}[t]
\centering

\includegraphics[width=0.45\textwidth]{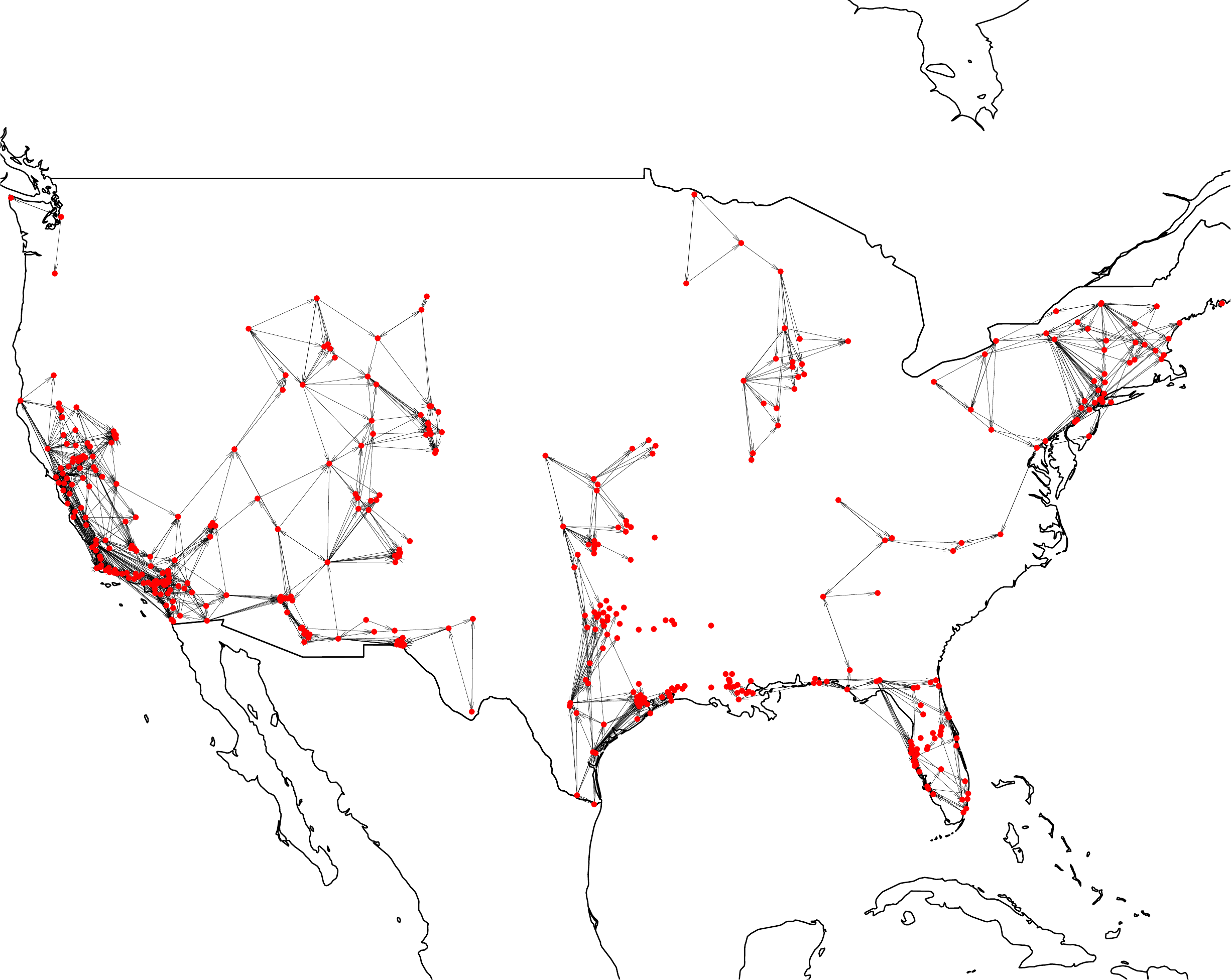}
\hspace{1cm}
\includegraphics[width=0.45\textwidth]{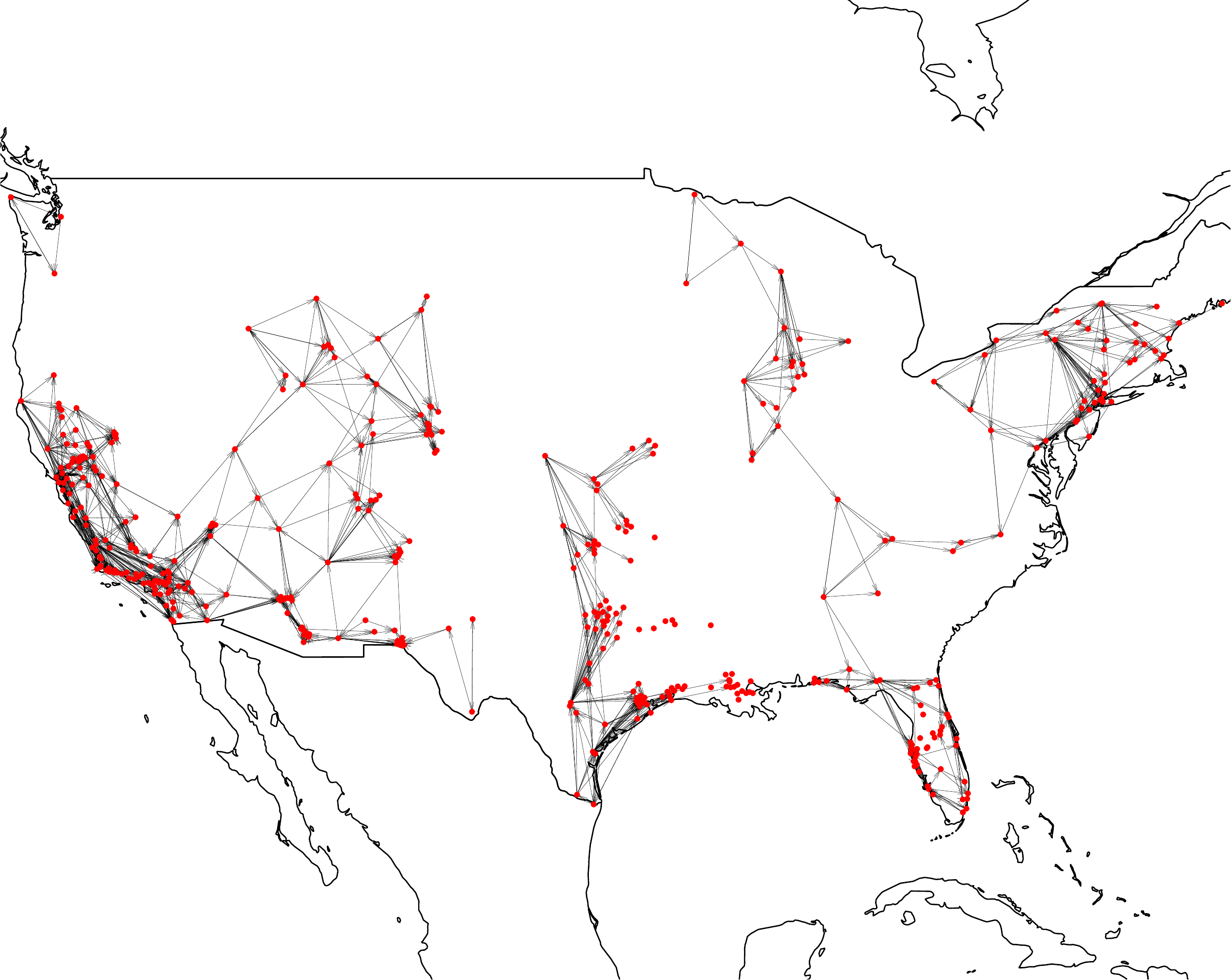}
\caption{Air pollution in the U.S.A.: autoregressive coefficients estimated by the two-step $\ell_1$-penalized least squares method with estimate $\wm = 239$ (left) and upper bound $\rho = 250$ (right),
where the upper bound is based on scientific evidence.
Monitors are connected by edges if the estimates of the corresponding autoregressive coefficients are non-zero.
The results demonstrate that the two-step $\ell_1$-penalized least squares method respects the fact that 24-hour dependence is local.}
\label{comparison2}
\end{figure}
}

We compare the two-step $\ell_1$-penalized least squares method with an estimate $\wm$ of $\rho$ to the two-step $\ell_1$-penalized least squares method with an upper bound on $\rho$ given by $\rho = 250$ and the $\ell_1$-penalized least squares method of \citet{BaMi13}.
The upper bound $\rho = 250$ is based on scientific evidence \citep{Raetal97},
which suggests that $\rho \leq 250$.
\hide{
The use of upper bounds on $\rho$,
if available,
is justified by Theorem \ref{theorem.known}.
}
The $\ell_1$-penalized least squares method of \citet{BaMi13} is the most attractive high-dimensional method available,
as discussed in Remark \ref{com.1} in Section \ref{sec:theory}.

The two-step $\ell_1$-penalized least squares method estimates $\rho$ by $\wm = 239$.
It is more than 8 times faster than the $\ell_1$-penalized least squares method and reduces the  out-of-sample 24-hour ahead forecast mean squared error by $4\%$.
If the upper bound $\rho = 250$ is used and hence $\rho$ is not estimated,
the two-step $\ell_1$-penalized least squares method is more than 18 times faster than the $\ell_1$-penalized least squares method.

The graphs estimated by the $\ell_1$-penalized least squares method and the two-step $\ell_1$-penalized least squares method with estimate $\wm = 239$ and upper bound $\rho = 250$ are shown in Figures \ref{comparison1} and \ref{comparison2},
respectively.
It is striking that the $\ell_1$-penalized least squares method reports a number of long-distance edges---some of them between monitors separated by more than 2,166 miles.
The long-distance edges conflict with scientific evidence,
which suggests that dependence local and that $\rho \leq 250$ \citep[e.g.,][]{Raetal97}:
it is not believed that today's Ozone levels on the East Coast can directly affect tomorrow's Ozone levels on the West Coast,
because Ozone cannot travel long distances \citep[see, e.g.,][]{Raetal97}.
In contrast, 
the two-step $\ell_1$-penalized least squares method reports that the estimated range of 24-hour dependence is $\wm = 239$,
which is consistent with scientific evidence \citep[e.g.,][]{Raetal97}.

\subsection{Zooming in: pollution in the Gulf region}
\label{app2}

\ghoster{
\begin{figure}[t]
\begin{center}
\includegraphics[width=0.7\textwidth]{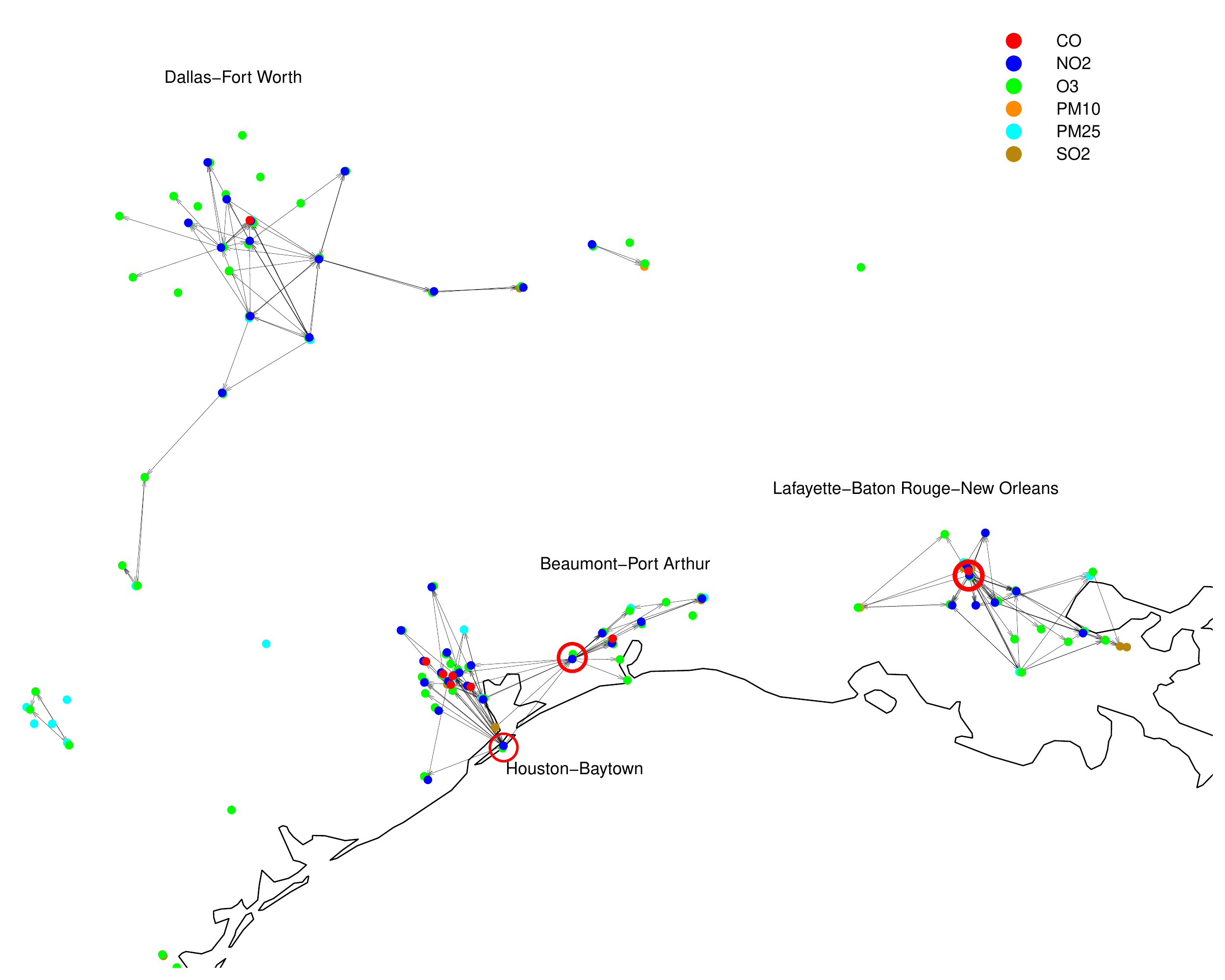}
\caption{
\label{fig:gulf.model}
Air pollution in the Gulf of Mexico region: autoregressive coefficients estimated by the two-step $\ell_1$-penalized least squares method from daily measurements of 6 pollutants.
Monitors are connected by edges if the estimates of the corresponding autoregressive coefficients are non-zero. 
Monitors with at least 18 outgoing edges are indicated by red circles.
}
\end{center}
\end{figure}
}

We zoom in on the Gulf of Mexico region and consider the 24-hour transport of $6$ pollutants:
Ozone ($O_3$), particle matter ($PM10$ and $PM2.5$), Carbon monoxide ($CO$), Nitrogen dioxide ($NO_2$), and Sulfur dioxide ($SO_2$).
The data set consists of $N =$ 1,826 observations of the 6 pollutants recorded by $k =$ 199 monitors between January 2010 and December 2014.
45.2\% of the time series are Ozone time series, 22.6\% are $NO_2$, 15.6\% are $PM25$, 9.1\% are $SO_2$, 5.5\% are $CO$, and 2.0\% are $PM10$.
\hide{
Some monitors keep track of multiple pollutants.
We treat such monitors as multiple monitors,
because we want to attribute 
e.g.,
we treat a monitor that keeps track of both Ozone and Nitrogen dioxide as two separate monitors with the same location.
Observe that treating such monitors as multiple monitors does not affect the radius $\rho$,
because $\rho$ is defined as the maximum distance of pairs of nodes with an edge,
but it increases the number of time series $k$ and thus the number of 
}

We estimate the model by the two-step $\ell_1$-penalized least squares method.
In Step 1,
we ensure that monitors of all 6 pollutants are well-represented in the sample by generating a stratified sample of size 20,
where the sample size of monitors of a pollutant is proportional to the total number of monitors of the pollutant in the Gulf of Mexico region.
The graph estimated by the two-step $\ell_1$-penalized least squares method is presented in Figure \ref{fig:gulf.model}.
Most edges are $NO_2 \rightarrow NO_2$ edges,
while most cross-pollutant edges are $NO_2 \rightarrow O_3$ edges,
which may be due to the chemical reaction that transforms Nitrogen oxides into Ozone in the presence of sunlight. 

There are two eye-catching facts in Figure \ref{fig:gulf.model}.
First,
there are $4$ clusters,
Dallas---Fort Worth, Houston---Baytown, Beaumont---Port Arthur, and Lafayette---Baton Rouge---New Orleans,
corresponding to industrial and metropolitan areas in the Gulf of Mexico region.
Second,
while the dependence structure is sparse and the median number of outgoing edges of monitors is 1, 
there are 3 monitors with at least 18 outgoing edges,
most of which are positive.
The large number of positive outgoing edges---i.e., positive autoregressive coefficients---suggests that pollution at those 3 locations tends to drive up pollution in neighboring regions.
It turns out that all of them are home to large industrial complexes,
including some of the largest oil refineries in the U.S.A.
These findings suggest that neighboring regions have reason to be concerned with the activities of the industrial sectors in those areas. 

\section{Discussion} 
\label{sec:discussion}

In practice, 
the two-step $\ell_1$-penalized least squares method can be used to decompose high-\linebreak
dimensional multivariate time series into lower-dimensional multivariate time series.
These lower-dimensional multivariate time series can then be studied in more detail by other methods \citep[see, e.g.,][]{CrWi11}.

There are multiple extensions of the two-step $\ell_1$-penalized least squares method that would be interesting to explore in the future.

One interesting extension would be to impose a parametric form on the transition matrices $\bA_1$, $\dots$, $\bA_L$ and the variance-covariance matrix $\bSigma$, 
i.e.,
to allow $\bA_1, \dots, \bA_L$ and $\bSigma$ to depend on distance in some parametric form.
To do so would require additional model assumptions,
but it could reduce statistical error.

A second interesting extension would be to assume that the radius of the past-present dependence captured by $\bA_1, \dots, \bA_L$ may not be the same as the radius of the present-present dependence captured by $\bSigma$.
Such extensions would make sense in applications where the present-present dependence captured by $\bSigma$ is more local than the past-present dependence captured by $\bA_1, \dots, \bA_L$.

Last,
but not least,
the two-step $\ell_1$-penalized least squares method is not restricted to high-dimensional vector autoregressive processes.
It can be extended to other high-dimensional models,
such as high-dimensional regression models and high-dimensional graphical models \citep[e.g.,][]{MeBu06,RaWaLa10},
as long as additional structure of the form considered here is available and consistent model selection in high dimensions is possible.

A potential problem in applications---as in other applications of multivariate statistics---are omitted variables.
The two-step $\ell_1$-penalized least squares method can be affected by omitted variables in both the first and second step of the method.
In the first step,
where sampled nodes are regressed on all other nodes,
excluded nodes can give rise to false-positive long-distance edges that produce large overestimates of the radius.
Large overestimates of the radius can increase computing time and statistical error in the second step,
where all nodes are regressed on all other nodes within the estimated radius.
Thus,
investigators should avoid omitted variable problems whenever possible.

\ghost{
\section*{Acknowledgements}

The research of the first two authors was supported by NSF award DMS-1513644. 
}

\section*{Supplementary material} 

All theoretical results are proved in the Appendix.
The data and the {\tt R} source code we used in Sections \ref{sec:sim} and \ref{sec:application} are contained in the supplementary archive.


\bibliographystyle{asa1}

\bibliography{sergii}

\ghost{
\newpage

\setcounter{page}{1}

\hide{
\begin{center}
\LARGE
{Supplementary Material:

High-Dimensional Multivariate Time Series With Additional Structure}\s\s
\\

\normalsize
{\large Michael Schweinberger\hspace{2cm} Sergii Babkin\hspace{2cm} Katherine Ensor\footnote{Address: Department of Statistics, Rice University, 6100 Main St, Houston, Texas 77005. Email: \email{michael.schweinberger@rice.edu}.}}
\hide{
\s
\\
{\em Department of Statistics, Rice University, 6100 Main St, Houston, Texas 77005, U.S.A.}\s
\\
\email{michael.schweinberger@rice.edu} \email{babkin.sergii@rice.edu} \email{ensor@rice.edu}
}
\end{center}
}

\begin{appendix}

\section{Proofs}

We prove Theorems \ref{theorem.known} and \ref{theorem.unknown} in Appendices \ref{appendix1} and \ref{appendix2},
respectively.

\subsection{Proof of Theorem \ref{theorem.known}}
\label{appendix1}

Let $\delta \geq 0$.
It is convenient to express the estimator $\wh\bbeta_{[0, \rho-\delta]}$ of $\bbeta_{[0,\rho-\delta]}^\star$ obtained in Step 2 of the two-step $\ell_1$-penalized least squares method as the solution of the $M$-estimation problem
\beno
\label{local.optimization}
\wh\bbeta_{[0, \rho-\delta]}
\in \argmin\limits_{\bbeta_{[0, \rho-\delta]}}\left[- 2\, \bbeta_{[0, \rho-\delta]}^\top\, \wh\bgamma_{[0, \rho-\delta]} + \bbeta_{[0, \rho-\delta]}^\top\, \wh\bGamma_{[0, \rho-\delta], [0, \rho-\delta]}\, \bbeta_{[0, \rho-\delta]} + \lambda_{2}\, \norm{\bbeta_{[0, \rho-\delta]}}_1\right].
\ee
We need three lemmas to prove Theorem \ref{theorem.known}.

\lemma
\label{lemma:reduced_model1}
Assume $N \geq c_0\, s \log p$ ($c_0 > 1$).
Then,
for all $\delta \geq 0$,
with at least probability
$1 - 2\, \exp(- c_1\, N)$,
\be
\label{condition1}
\bb^{\top}\; \wh\bGamma_{[0, \rho-\delta], [0, \rho-\delta]}\; \bb 
\;\;\geq\;\; \alpha\, \norm{\bb}_2^2 - \tau\, \norm{\bb}^2_1\;\; \mbox{ for all }\; \bb\; \in\; \mathbb{R}^{p(0,\rho-\delta)}.
\ee

\begin{proof}
Observe that $\wh\bGamma_{[0, \rho-\delta], [0, \rho-\delta]}$ can be written as $\wh\bGamma_{[0, \rho-\delta], [0, \rho-\delta]} = \bD^\top\, \wh\bGamma\, \bD$,
where $\bD$ is a 0-1 elimination matrix of suitable order that eliminates the elements of $\wh\bGamma$ that are not elements of $\wh\bGamma_{[0, \rho-\delta], [0, \rho-\delta]}$.
By Assumption \ref{ass1},
for all $\bb \in \mathbb{R}^{p(0,\rho-\delta)}$,
\be
\label{c1.local}
\bb^{\top}\, \wh\bGamma_{[0, \rho-\delta], [0, \rho-\delta]}\, \bb
\;=\; (\bE\, \bb)^{\top}\, \wh\bGamma\, (\bE\, \bb)
\;\geq\; \alpha\, \norm{\bE\, \bb}_2^2 - \tau\, \norm{\bE\, \bb}_1^2
\;=\; \alpha\, \norm{\bb}_2^2 - \tau\, \norm{\bb}_1^2,
\ee
where $||\bE\, \bb||_i = ||\bb||_i$, $i = 1, 2$, 
because the $p$-vector $\bE\, \bb$ consists of the $p(0,\rho-\delta)$ elements of $\bb$ and $p-p(0,\rho-\delta)$ $0$'s.
\hide{
which implies $||\bE\, \bb||_1 = ||\bb||_1$ and $||\bE\, \bb||_2 = ||\bb||_2$,
}
The lower bound \eqref{c1.local} holds as long as Assumption \ref{ass1} holds.
By Proposition 4.2 of \citet{BaMi13},
the probability that Assumption \ref{ass1} is violated is bounded above by $2\, \exp(- c_1\, N)$ provided $N \geq c_0\, s \log p$ ($c_0 > 1$).
\end{proof}

\lemma
\label{lemma:reduced_model2}
Assume $N \geq \log p(0,\rho-\delta)$.
Then,
for all $\delta \geq 0$,
with at least probability
$1 - 6\, \exp\left(- c_2 \log p(0,\rho-\delta)\right)$,
\be
\label{condition2}
\norm{\wh\bgamma_{[0, \rho-\delta]} - \wh\bGamma_{[0, \rho-\delta], [0, \rho-\delta]}\, \bbeta^\star_{[0, \rho-\delta]}}_\infty
&\leq& \mQ\, \sqrt{\dfrac{\log p(0,\rho-\delta)}{N}}.
\ee

\begin{proof}
The proof proceeds along the lines of Proposition 4.3 of \citet[][supplement, pp.\ 6--7]{BaMi13} by applying concentration inequality (2.11) of \citet{BaMi13} to bound the probability of 
\beno
\norm{\wh\bgamma_{[0, \rho-\delta]} - \wh\bGamma_{[0, \rho-\delta], [0, \rho-\delta]}\, \bbeta^\star_{[0, \rho-\delta]}}_\infty
&>& 2\, \pi\, \dfrac{\mQ}{a}\, \eta,
\ee
where $a > 0$ and $\eta > 0$.
Choosing $\eta = (a / (2\, \pi))\, \sqrt{\log p(0,\rho-\delta) / N}$ gives
\be
\label{bound}
\norm{\wh\bgamma_{[0, \rho-\delta]} - \wh\bGamma_{[0, \rho-\delta], [0, \rho-\delta]}\, \bbeta^\star_{[0, \rho-\delta]}}_\infty 
&>& \mQ\, \sqrt{\dfrac{\log p(0,\rho-\delta)}{N}}.
\ee
The concentration inequality (2.11) of \citet{BaMi13} and a union bound show that,
provided $N \geq \log p(0,\rho-\delta)$,
the probability of \eqref{bound} is bounded above by
\beno
6\, \exp\left(- c\, N\, \min\left(\eta, \eta^2\right) + \log p(0,\rho-\delta)\right)
&\leq& 6\, \exp\left(- c_2 \log p(0,\rho-\delta)\right).
\ee
\end{proof}

\hide{
The concentration result (2.11) of \citet{BaMi13} along with $N \geq \log p(0,\rho-\delta)$ and a union bound shows that there exists a constant $c_2 > 0$ such that the probability of event \eqref{bound} is bounded above by
\beno
&& 6\, \exp\left[- c\, N\, \min\left(\dfrac{a}{2\, \pi}\, \sqrt{\dfrac{\log p(0,\rho-\delta)}{N}}, \dfrac{a^2}{4\, \pi^2}\, \dfrac{\log p(0,\rho-\delta)}{N}\right) + \log p(0,\rho-\delta)\right]\s
\\
&=& 6\, \exp\left[- c\, \min\left(\dfrac{a}{2\, \pi}\, \sqrt{N \log p(0,\rho-\delta)}, \dfrac{a^2}{4\, \pi^2}\, \log p(0,\rho-\delta)\right) + \log p(0,\rho-\delta)\right]\s
\\
&\leq& 
\hide{
6\, \exp\left[- c\, \min\left(\dfrac{a}{2\, \pi}\, \sqrt{\log p(0,\rho-\delta) \log p(0,\rho-\delta)}, \dfrac{a^2}{4\, \pi^2}\, \log p(0,\rho-\delta))\right) + \log p(0,\rho-\delta)\right]\s
\\
&=&
6\, \exp\left[- c\, \min\left(\dfrac{a}{2\, \pi}\, \log p(0,\rho-\delta), \dfrac{a^2}{4\, \pi^2}\, \log p(0,\rho-\delta)\right) + \log p(0,\rho-\delta)\right]\s
\\
&=&
}
6\, \exp\left[- c\, \min\left(\dfrac{a}{2\, \pi}, \dfrac{a^2}{4\, \pi^2}\right) \log p(0,\rho-\delta) + \log p(0,\rho-\delta)\right]\s
\\
&=& 6\, \exp\left[- \left(c\, \min\left(\dfrac{a}{2\, \pi}, \dfrac{a^2}{4\, \pi^2}\right) - 1\right) \log p(0,\rho-\delta)\right]\s
\\
&=& 6\, \exp\left(- c_2 \log p(0,\rho-\delta)\right),
\ee
where the constant $a > 0$,
which is defined in condition Assumption \ref{ass2} and which we are free to choose,
is chosen so that
\beno
c_2
&=& c\, \min\left(\dfrac{a}{2\, \pi}, \dfrac{a^2}{4\, \pi^2}\right) - 1
&>& 0.
\ee
}

\lemma
\label{lemma.bound}
Assume that conditions \eqref{condition1} and \eqref{condition2} are satisfied and $\lambda_2 \geq 4\, \mQ\, \sqrt{\log p(0,\rho-\delta) / N}$.
Then,
for all $\delta \geq 0$,
\beno
\norm{\wh\bbeta_{[0, \rho-\delta]} - \bbeta^\star_{[0, \rho-\delta]}}_2
&\leq& \dfrac{16\, \sqrt{s}\, \lambda_{2}}{\alpha}.
\ee

\begin{proof}
By definition of $\wh\bbeta_{[0, \rho-\delta]}$,
for all $\bbeta_{[0, \rho-\delta]} \in \mR^{p(0,\rho-\delta)}$,
\be
\label{inequality00}
&& -2\, \wh\bbeta_{[0, \rho-\delta]}^\top\, \wh\bgamma_{[0, \rho-\delta]} + \wh\bbeta_{[0, \rho-\delta]}^\top\, \wh\bGamma_{[0, \rho-\delta], [0, \rho-\delta]}\, \wh\bbeta_{[0, \rho-\delta]}
+ \lambda_{2}\, \norm{\wh\bbeta_{[0, \rho-\delta]}}_1\s
\\
&\leq& -2\, \bbeta_{[0, \rho-\delta]}^\top\, \wh\bgamma_{[0, \rho-\delta]} + \bbeta_{[0, \rho-\delta]}^\top\, \wh\bGamma_{[0, \rho-\delta], [0, \rho-\delta]}\, \bbeta_{[0, \rho-\delta]}
+ \lambda_{2}\, \norm{\bbeta_{[0, \rho-\delta]}}_1.
\ee
Set $\bbeta_{[0, \rho-\delta]} = \bbeta_{[0, \rho-\delta]}^\star$ and $\bv = \wh\bbeta_{[0, \rho-\delta]} - \bbeta_{[0, \rho-\delta]}^\star$.
Then \eqref{inequality00} reduces to
\be
\label{inequality01}
\bv^\top\, \wh\bGamma_{[0, \rho-\delta], [0, \rho-\delta]}\, \bv\s
\\
\leq 2\, \bv^\top (\wh\bgamma_{[0, \rho-\delta]} - \wh\bGamma_{[0, \rho-\delta], [0, \rho-\delta]}\, \bbeta^\star_{[0, \rho-\delta]}) + \lambda_{2}\, (\norm{\bbeta^\star_{[0, \rho-\delta]}}_1 - \norm{\bbeta^\star_{[0, \rho-\delta]} - \bv}_1).
\ee
The first term on the right-hand side of \eqref{inequality01} can be bounded by using condition \eqref{condition2} and $\lambda_2 \geq 4\, \mQ\, \sqrt{\log p(0,\rho-\delta) / N}$:
\be
\label{firstterm}
2\, \bv^\top (\wh\bgamma_{[0, \rho-\delta]} - \wh\bGamma_{[0, \rho-\delta], [0, \rho-\delta]}\, \bbeta^\star_{[0, \rho-\delta]})
\hide{
\\
&\leq& 2\, \norm{\bv}_1\, \norm{\wh\bgamma_{[0, \rho-\delta]} - \wh\bGamma_{[0, \rho-\delta], [0, \rho-\delta]}\, \bbeta^\star_{[0, \rho-\delta]}}_\infty\s
\\
&\leq& 2\, \norm{\bv}_1\, \mQ\, \sqrt{\dfrac{\log p(0,\rho-\delta)}{N}}\s
\\
\hide{
&\leq& 2\, \norm{\bv}_1\, \mQ\, \sqrt{\dfrac{\log p(0,\wm)}{N}}\s
\\
}
}
\;\leq\; \dfrac{\lambda_{2}}{2}\, \norm{\bv}_1
\;=\; \dfrac{\lambda_{2}}{2}\, (\norm{\bvs}_1 + \norm{\bvc}_1),
\ee
where $\bvs$ and $\bvc$ are the subvectors of $\bv$ corresponding to the support $\supp[0,\rho-\delta]$ of $\bbeta_{[0, \rho-\delta]}^\star$ and its complement $\overline\supp[0,\rho-\delta]$,
respectively.
The second term on the right-hand side of \eqref{inequality01} can be bounded as follows:
\be
\label{secondterm}
\lambda_{2}\, (\norm{\bbeta^\star_{[0, \rho-\delta]}}_1 - \norm{\bbeta^\star_{[0, \rho-\delta]} - \bv}_1)
&\leq& \lambda_{2}\, (\norm{\bvs}_1 - \norm{\bvc}_1)
\ee
using the triangle inequality
\beno
\norm{\bbeta^\star_{[0, \rho-\delta]}}_1
&=& \norm{\bbeta^\star_{\mbS[0, \rho-\delta]}}_1
\hide{
&\leq& \norm{\bbeta^\star_{\mbS[0, \rho-\delta]} - \bvs}_1 + \norm{\bvs - \bm{0}_s}_1\s
\\
&&}
&\leq& \norm{\bbeta^\star_{\mbS[0, \rho-\delta]} - \bvs}_1 + \norm{\bvs}_1.
\ee
\hide{
along with
\be
\norm{\bbeta^\star_{[0, \rho-\delta]} - \bv}_1
\hide{
&=& \norm{\bbeta^\star_{\mbS[0, \rho-\delta]} - \bvs}_1 + \norm{\bbeta^\star_{\overline\mbS[0, \rho-\delta]} - \bvc}_1\s
\\
}
&=& \norm{\bbeta^\star_{\mbS[0, \rho-\delta]} - \bvs}_1 + \norm{\bvc}_1.
\ee
}
\hide{
implying
\be
\norm{\bbeta^\star_{\mbS[0, \rho-\delta]} - \bvs}_1 - \norm{\bbeta^\star_{[0, \rho-\delta]} - \bv}_1
&\leq& \norm{\bbeta^\star_{\mbS[0, \rho-\delta]} - \bvs}_1 + \norm{\bvs}_1 - \norm{\bbeta^\star_{\mbS[0, \rho-\delta]} - \bvs}_1 - \norm{\bvc}_1\s
\\
&\leq& \norm{\bvs}_1 - \norm{\bvc}_1.
\ee
}
Therefore,
combining \eqref{inequality01} with \eqref{firstterm} and \eqref{secondterm},
\be
\label{basic}
0
&\leq& \bv^\top\, \wh\bGamma_{[0, \rho-\delta], [0, \rho-\delta]}\, \bv
\hide{
\s
\\
\hide{
&\leq& 2\, \bv^\top (\wh\bgamma_{[0, \rho-\delta]} - \wh\bGamma_{[0, \rho-\delta], [0, \rho-\delta]}\, \bbeta^\star_{[0, \rho-\delta]}) + \lambda_{2} (\norm{\bbeta^\star_{[0, \rho-\delta]}}_1 - \norm{\bbeta^\star_{[0, \rho-\delta]} - \bv}_1)\s\s
\\
}
&\leq& \dfrac{\lambda_{2}}{2}\, (\norm{\bvs}_1 + \norm{\bvc}_1) + \lambda_{2}\, (\norm{\bvs}_1 - \norm{\bvc}_1)\s\s
\\
&=& 
}
&\leq& \dfrac{3\,\lambda_{2}}{2} \norm{\bvs}_1 - \dfrac{\lambda_{2}}{2}\, \norm{\bvc}_1.
\ee
Thus,
$\norm{\bvc}_1 \leq 3\, \norm{\bvs}_1$,
implying
\be
\label{equat}
\norm{\bv}_1
&=& \norm{\bvs}_1 + \norm{\bvc}_1
&\leq& 4\, \norm{\bvs}_1
\hide{
&\leq& 4\, \sqrt{s}\, \norm{\bvs}_2
}
&\leq& 4\, \sqrt{s}\, \norm{\bv}_2.
\ee
An upper bound on $\bv^\top\, \wh\bGamma_{[0, \rho-\delta], [0, \rho-\delta]}\,\bv$ can therefore be obtained by using \eqref{basic} and \eqref{equat}: 
\beno
\hide{
0 &\leq&
}
\bv^\top\, \wh\bGamma_{[0, \rho-\delta], [0, \rho-\delta]}\, \bv
&\leq& \dfrac{3\,\lambda_{2}}{2} \norm{\bvs}_1 - \dfrac{\lambda_{2}}{2}\, \norm{\bvc}_1
\hide{
&\leq& \dfrac{3\,\lambda_{2}}{2}\, \norm{\bvs}_1\s
\\
&\leq& \dfrac{3\, \lambda_{2}}{2}\, \norm{\bv}_1
}
&\leq& 2\, \lambda_{2}\, \norm{\bv}_1,
\ee
implying
\beno
\dfrac12\, \bv^\top\, \wh\bGamma_{[0, \rho-\delta], [0, \rho-\delta]}\, \bv
&\leq& \lambda_{2}\, \norm{\bv}_1
&\leq& 4\, \sqrt{s}\, \lambda_{2}\, \norm{\bv}_2.
\ee
A lower bound on $\bv^\top\, \wh\bGamma_{[0, \rho-\delta], [0, \rho-\delta]}\,\bv$ can be derived by using Lemma \ref{lemma:reduced_model1} and \eqref{equat} along with $s\, \tau \leq \alpha / 32$,
giving
\beno
\bv^\top\, \wh\bGamma_{[0, \rho-\delta], [0, \rho-\delta]}\, \bv
\;\geq\; \alpha\, \norm{\bv}_2^2 - \tau\, \norm{\bv}^2_1\s
\;\geq\; \alpha\, \norm{\bv}_2^2 - \tau\, 16\, s\, \norm{\bv}_2^2
\;\geq\; \dfrac{\alpha}{2}\, \norm{\bv}_2^2.
\ee
Combining the upper and lower bounds on $\bv^\top\, \wh\bGamma_{[0, \rho-\delta], [0, \rho-\delta]}\, \bv$ gives
\beno
\dfrac{\alpha}{4}\, \norm{\bv}_2^2
&\leq& \dfrac{1}{2}\, \bv^\top\, \wh\bGamma_{[0, \rho-\delta], [0, \rho-\delta]}\, \bv
&\leq& 4\, \sqrt{s}\, \lambda_{2}\, \norm{\bv}_2,
\ee
implying
\beno
\norm{\bv}_2
&=& \norm{\wh\bbeta_{[0, \rho-\delta]} - \bbeta^\star_{[0, \rho-\delta]}}_2
&\leq& \dfrac{16\, \sqrt{s}\, \lambda_{2}}{\alpha}.
\ee
\hide{
It can be shown that,
by using conditions \eqref{condition1} and \eqref{condition2} along with $\lambda_2 \geq 4\, \mQ\, \sqrt{\log p(0,\wm) / N}$,
$\bv^\top\, \wh\bGamma_{[0, \rho-\delta], [0, \rho-\delta]}\,\bv$ can be bounded as follows:
\beno
\dfrac{\alpha}{4}\, \norm{\bv}_2^2
&\leq& \dfrac{1}{2}\, \bv^\top\, \wh\bGamma_{[0, \rho-\delta], [0, \rho-\delta]}\, \bv
&\leq& 4\, \sqrt{s}\, \lambda_{2}\, \norm{\bv}_2,
\ee
implying
\beno
\norm{\bv}_2
&=& \norm{\wh\bbeta_{[0, \rho-\delta]} - \bbeta^\star_{[0, \rho-\delta]}}_2
&\leq& \dfrac{16\, \sqrt{s}\, \lambda_{2}}{\alpha}.
\ee
}
\end{proof}

\begin{proof}
{\em Theorem \ref{theorem.known}.}
By Lemma \ref{lemma.bound} with $\delta = 0$,
as long as conditions \eqref{condition1} and \eqref{condition2} are satisfied,
\be
\label{mybound0}
\norm{\wh\bbeta - \bbeta^\star}_2
&=& \norm{\wh\bbeta_{[0, \rho]} - \bbeta^\star_{[0, \rho]}}_2 
&\leq& \dfrac{16\, \sqrt{s}\, \lambda_{2}}{\alpha},
\ee
where we used the fact that all elements of $\wh\bbeta$ and $\bbeta^\star$ corresponding to edges at distances $d > \rho$ are $0$.
The upper bound \eqref{mybound0} holds as long as conditions \eqref{condition1} and \eqref{condition2} hold. 
By Lemmas \ref{lemma:reduced_model1} and \ref{lemma:reduced_model2} with $\delta = 0$ along with $N \geq c_0\, s \log p \geq \log p(0,\rho)$ ($c_0 > 1$) and a union bound,
the probability that \eqref{condition1} or \eqref{condition2} are violated is bounded above by
\beno
2\, \exp(- c_1\, N) + 6\, \exp(- c_2 \log p(0,\rho)).
\ee
\end{proof}

\subsection{Proof of Theorem \ref{theorem.unknown}}
\label{appendix2}

We need three additional lemmas to prove Theorem \ref{theorem.unknown}.

\lemma
\label{proposition1}
For all $\delta > 0$,
the probability that none of the nodes $i \in \mS(\delta)$ is sampled is bounded above by
\beno
\exp\left(-\dsum_{i\in\mS(\delta)} \theta_i\right).
\ee

\begin{proof}
By definition of $\rho > 0$,
for all $\delta > 0$, 
there exists at least one node with incoming edges at distances $d \in [\rho-\delta, \rho]$,
thus $\mS(\delta)$ is non-empty.
Since nodes $i$ are sampled independently with probabilities $0 < \theta_i < 1$,
the probability that none of the nodes $i \in \mS(\delta)$ is sampled is bounded above by
\beno
\exp\left(\dsum_{i\in\mS(\delta)} \log(1 - \theta_i)\right)
&\leq& \exp\left(-\dsum_{i\in\mS(\delta)} \theta_i\right).
\ee
\end{proof}

\lemma
\label{proposition2}
Let $\beta_{\min}^\star \geq 32\, \sqrt{s}\, \lambda_{1} / \alpha$,
where $\lambda_1 \geq 4\, \mQ\, \sqrt{\log p / N}$.
Then,
for any $\delta > 0$ and any non-empty subset $\mA \subseteq \mS(\delta)$,
the probability that none of the incoming edges of nodes $i \in \mA$ at distances $d \in [\rho-\delta, \rho]$ is detected is bounded above by
\beno
2\, \exp(- c_1\, N) + 6\, \exp(- c_2 \log p).
\ee

\begin{proof}
By definition of $\rho > 0$,
for all $\delta > 0$, 
there exists at least one node with incoming edges at distances $d \in [\rho-\delta, \rho]$,
thus $\mS(\delta)$ is non-empty.
Consider any non-empty subset $\mA \subseteq \mS(\delta)$.
Let $\mG$ be the event that all incoming edges of all nodes $i \in \mA$ are detected and $\mB$ be its complement.
Then the event that none of the incoming edges of nodes $i \in \mA$ at distances $d \in [\rho-\delta, \rho]$ is detected is contained in $\mB$ and the probability of the event of interest is bounded above by the probability of $\mB$.
To bound the probability of $\mB$,
let $\wh\bbeta_{\mN}$ and $\wh\bbeta_{\mA}$ be solutions of optimization problems \eqref{optimization1a} and \eqref{optimization1b},
respectively,
and observe that $\mG$ is implied by
\beno
\dfrac{2}{\beta_{\min}^\star}\, \norm{\wh\bbeta_{\mA} - \bbeta_{\mA}^\star}_\infty
&\leq& 1.
\ee
Since
\beno
\dfrac{2}{\beta_{\min}^\star}\, \norm{\wh\bbeta_{\mA} - \bbeta_{\mA}^\star}_\infty
&\leq& \dfrac{2}{\beta_{\min}^\star}\, \norm{\wh\bbeta_{\mN} - \bbeta_{\mN}^\star}_\infty,
\ee
we have,
by Assumptions \ref{ass1} and \ref{ass2} and $\beta_{\min}^\star \geq 32\, \sqrt{s}\, \lambda_{1} / \alpha$,
\be
\label{firstbound}
\dfrac{2}{\beta_{\min}^\star}\, \norm{\wh\bbeta_{\mN} - \bbeta_{\mN}^\star}_\infty
&\leq& \dfrac{2}{\beta_{\min}^\star}\, \norm{\wh\bbeta_{\mN} - \bbeta_{\mN}^\star}_2
&\leq& \dfrac{2}{\beta_{\min}^\star}\, \dfrac{16\, \sqrt{s}\, \lambda_{1}}{\alpha}
&\leq& 1.
\ee
The bound $||\wh\bbeta_{\mN} - \bbeta_{\mN}^\star||_2 \leq 16\, \sqrt{s}\, \lambda_{1} / \alpha$ used in \eqref{firstbound} follows from Proposition 4.1 of \citet[][]{BaMi13} and holds as long as Assumptions \ref{ass1} and \ref{ass2} hold.
Therefore, 
$\mG$ occurs as long as both Assumptions \ref{ass1} and \ref{ass2} hold,
whereas $\mB$ occurs when either Assumption \ref{ass1} or Assumption \ref{ass2} or both are violated.
A union bound along with $N \geq c_0\, s \log p \geq \log p$ ($c_0 > 1$) shows that the probability of $\mB$,
and thus the event of interest, 
is bounded above by
\be
\label{union}
\hide{
&=& \mbP(\mC_1 \cup \mC_2)
&\leq& \mbP(\mC_1) + \mbP(\mC_2)
}
2\, \exp(- c_1\, N) + 6\, \exp(- c_2 \log p),
\ee
where the two terms in \eqref{union} are upper bounds on the probabilities that Assumption \ref{ass1} or Assumption \ref{ass2} are violated,
which follow from Propositions 4.2 and 4.3 of \citet{BaMi13},
respectively.
\end{proof}

\lemma
\label{thirdlemma}
Consider $N \geq c_0\, s \log p$ ($c_0 > 1$) observations from a stable $L$-th order vector autoregressive process with radius $\rho > 0$.
Assume that components $i$ are sampled independently with probabilities $0 < \theta_i < 1$,
the minimum signal strength is $\beta_{\min}^\star = \min_{i \in \supp} |\beta_i^\star| \geq 32\, \sqrt{s}\, \lambda_1 / \alpha > 0$,
and the regularization parameter $\lambda_{1}$ satisfies
\beno
\label{lambda1}
\lambda_{1}
&\geq& 4\, \mQ\, \sqrt{\dfrac{\log p}{N}}.
\ee
Then,
for all $\delta > 0$,
\beno
\mbP\left(\wm - \rho\, <\, -\delta\right)
&\leq& 2\, \exp(- c_1\, N) + 6\, \exp(- c_2 \log p) + \exp\left(-\dsum_{i\in\mS(\delta)} \theta_i\right).
\ee

\begin{proof}
By definition of $\rho > 0$,
for all $\delta > 0$,
there exists at least one node with incoming edges at distances $d \in [\rho-\delta, \rho]$,
thus $\mS(\delta)$ is non-empty.
Let $\mG_1$ be the event that at least one node $i \in \mS(\delta)$ with incoming edges at distances $d \in [\rho-\delta, \rho]$ is sampled and that at least one of its incoming edges at distances $d \in [\rho-\delta, \rho]$ is detected and $\mG_2$ be the event that at least one false-positive incoming edge of nodes $i\in\mS$ at distances $d \in [\rho-\delta, \infty)$ is reported.
Then the event $\wm-\rho \geq -\delta$ is equivalent to the event $\mG_1 \cup \mG_2$.
Thus,
the probability of event $\wm-\rho \geq -\delta$ is bounded below by
\beno
\mbP(\wm - \rho \geq -\delta)
\hide{
&=& \mbP((\mG_1 \cap \mG_2) \cup \mG_3)\s
\\
&\geq& \mbP(\mG_1 \cap \mG_2)\s
\\
&=& \mbP(\mG_1)\, \mbP(\mG_2 \mid \mG_1)\s
\\
}
&=& \mbP(\mG_1 \cup \mG_2)
\;\geq\; \mbP(\mG_1)\s
\hide{
&\geq& \left[1 - \exp\left(-\dsum_{i\in\mS(\delta)} \theta_i\right)\right]\, \left[1 - 2\, \exp(- c_1\, N) - 6\, \exp(- c_2 \log p)\right]\s
}
\hide{
\\
&=& 1 - 2\, \exp(- c_1\, N) - 6\, \exp(- c_2 \log p)\s
\\
&-& \exp\left(-\dsum_{i\in\mS(\delta)} \theta_i\right)\, \left[1 - 2\, \exp(- c_1\, N) - 6\, \exp(- c_2 \log p)\right]\s
\\
&=& 1 - 2\, \exp(- c_1\, N) - 6\, \exp(- c_2 \log p)\s
\\
&-& \exp\left(-\dsum_{i\in\mS(\delta)} \theta_i\right) + \exp\left(-\dsum_{i\in\mS(\delta)} \theta_i\right)\, \left[2\, \exp(- c_1\, N) + 6\, \exp(- c_2 \log p)\right]\s
}
\\
&\geq& 1 - 2\, \exp(- c_1\, N) - 6\, \exp(- c_2 \log p) - \exp\left(-\dsum_{i\in\mS(\delta)} \theta_i\right),
\ee
where we used a union bound along with Lemmas \ref{proposition1} and \ref{proposition2} to bound the probability of the complement of event $\mG_1$.
\end{proof}

\begin{proof}
{\em Theorem \ref{theorem.unknown}.}
For all $\delta > 0$,
\be
\label{t1}
&& \mbP\left(\norm{\wh\bbeta_{[0, \rho-\delta]} - \bbeta^\star_{[0, \rho-\delta]}}_2\, >\, \dfrac{16\, \sqrt{s}\, \lambda_{2}}{\alpha}\right)
\;\;\leq\;\; \mbP\left(\wm - \rho\, <\, -\delta\right)\s
\\
&+& \mbP\left(\left(\norm{\wh\bbeta_{[0, \rho-\delta]} - \bbeta^\star_{[0, \rho-\delta]}}_2\, >\, \dfrac{16\, \sqrt{s}\, \lambda_{2}}{\alpha}\right) \cap \Big(\wm - \rho \geq -\delta\Big)\right).
\ee
We bound the two terms on the right-hand side of \eqref{t1} one by one.

\s

{\em First term on the right-hand side of \eqref{t1}.}
By Lemma \ref{thirdlemma},
the first term on the right-hand side of \eqref{t1} is bounded above by 
\be
\label{rho.bound}
\mbP\left(\wm - \rho\, <\, -\delta\right)
&\leq& 2\, \exp(- c_1\, N) + 6\, \exp(- c_2 \log p) + \exp\left(-\dsum_{i\in\mS(\delta)} \theta_i\right).
\ee

\s

{\em Second term on the right-hand side of \eqref{t1}.}
We are interested in the intersection of the event that $\wm - \rho \geq -\delta$ and the event that
\be
\label{mybound}
\norm{\wh\bbeta_{[0, \rho-\delta]} - \bbeta^\star_{[0, \rho-\delta]}}_2
&>& \dfrac{16\, \sqrt{s}\, \lambda_{2}}{\alpha},
\ee
where $\lambda_2 \geq 4\, \mQ\, \sqrt{\log p(0,\rho-\delta) / N}$.
By Lemma \ref{lemma.bound} and 
$\lambda_2 \geq 4\, \mQ\, \sqrt{\log p(0,\rho-\delta) / N}$,
as long as conditions \eqref{condition1} and \eqref{condition2} are satisfied,
$\norm{\wh\bbeta_{[0, \rho-\delta]} - \bbeta^\star_{[0, \rho-\delta]}}_2$ is bounded above by
\be
\label{mybound}
\norm{\wh\bbeta_{[0, \rho-\delta]} - \bbeta^\star_{[0, \rho-\delta]}}_2
&\leq& \dfrac{16\, \sqrt{s}\, \lambda_{2}}{\alpha}.
\ee
By Lemmas \ref{lemma:reduced_model1} and \ref{lemma:reduced_model2} along with $N \geq c_0\, s \log p \geq \log p(0,\rho-\delta)$ ($c_0 > 1$) and a union bound,
the probability that \eqref{condition1} or \eqref{condition2} are violated is bounded above by
\be
\label{parameter.bound}
2\, \exp(- c_1\, N) + 6\, \exp(- c_2 \log p(0,\rho-\delta)).
\ee
Thus,
the second term on the right-hand side of \eqref{t1} is bounded above by \eqref{parameter.bound}.

\s

{\em Conclusion.}
Combining \eqref{t1}, \eqref{rho.bound}, and \eqref{parameter.bound} shows that 
\be
&& \mbP\left(\norm{\wh\bbeta_{[0, \rho-\delta]} - \bbeta^\star_{[0, \rho-\delta]}}_2\, >\, \dfrac{16\, \sqrt{s}\, \lambda_{2}}{\alpha}\right)\s
\\
&\leq& 2\, \exp(- c_1\, N) + 6\, \exp(- c_2 \log p) + \exp\left(-\dsum_{i\in\mS(\delta)} \theta_i\right)\s
\\
&+& 2\, \exp(- c_1\, N) + 6\, \exp(- c_2 \log p(0,\rho-\delta))\s
\\
&\leq& 4\, \exp(- c_1\, N) + 12\, \exp(- c_2 \log p(0,\rho-\delta)) + \exp\left(-\dsum_{i\in\mS(\delta)} \theta_i\right),
\ee
where we used the fact that $p(0,\rho-\delta) \leq p$ for all $\delta > 0$.
\end{proof}

\hide{
Along the same lines,
it can be shown that,
for any such $\delta > 0$, however small,
\beno
&& \mbP\left(\norm{\wh\bbeta - \bbeta^\star}_2\, >\, \dfrac{16}{\alpha} \mQ \sqrt{\dfrac{s \log p(0,\rho-\delta)}{N}} + \norm{\bbeta^\star_{(\rho-\delta, \rho]}}_2\right)\s
\\
&\leq& \mbP\left(\wm - \rho\, <\, -\delta\right)\s
\\
&+& \mbP\left(\norm{\wh\bbeta - \bbeta^\star}_2\, >\, \dfrac{16}{\alpha} \mQ \sqrt{\dfrac{s \log p(0,\rho-\delta)}{N}} + \norm{\bbeta^\star_{(\rho-\delta, \rho]}}_2\; \cap\; \wm - \rho \geq -\delta\right)\s
\\
&\leq& 4\, \exp(- c_1\, N) + 12\, \exp(- c_2 \log p(0,\rho-\delta)) + \exp\left(-\dsum_{i\in\mS(\delta)} \theta_i\right).
\ee
}

\hide{
\subsection{Proof of Theorem \ref{theorem.unknown}.2}

\hide{
Let $\delta > 0$ and $\wm - \rho \geq -\delta$.
It is convenient to express the estimator $\wh\bbeta_{[0, \rho-\delta]}$ of $\bbeta_{[0,\rho-\delta]}^\star$ obtained in Step 2 as the solution of the $M$-estimation problem
\beno
\label{local.optimization}
\wh\bbeta_{[0, \rho-\delta]}
\in \argmin\limits_{\bbeta_{[0, \rho-\delta]}}\left[- 2\, \bbeta_{[0, \rho-\delta]}^\top\, \wh\bgamma_{[0, \rho-\delta]} + \bbeta_{[0, \rho-\delta]}^\top\, \wh\bGamma_{[0, \rho-\delta], [0, \rho-\delta]}\, \bbeta_{[0, \rho-\delta]} + \lambda_{2}\, \norm{\bbeta_{[0, \rho-\delta]}}_1\right].
\ee
}

\begin{proof}
{\em Theorem \ref{theorem.unknown}.2.}
Consider $\wm < \rho$.
By Lemma \ref{lemma.bound},
as long as conditions \eqref{condition1} and \eqref{condition2} are satisfied,
\be
\label{mybound}
\norm{\wh\bbeta_{[0, \rho-\delta]} - \bbeta^\star_{[0, \rho-\delta]}}_2 
&\leq& \dfrac{16\, \sqrt{s}\, \lambda_{2}}{\alpha}.
\ee
By the triangle inequality,
\be
\label{mybound2}
\norm{\wh\bbeta - \bbeta^\star}_2
\hide{
&\leq& \norm{\wh\bbeta - \bbeta^\star_{0}}_2 + \norm{\bbeta^\star_{0} - \bbeta^\star}_2\s
\\
&\leq& \norm{\wh\bbeta_{[0,\rho-\delta]} - \bbeta^\star_{[0,\rho-\delta]}}_2 + \norm{\bbeta^\star_{(\rho-\delta, \rho]}}_2\s
\\
}
&\leq& \dfrac{16\, \sqrt{s}\, \lambda_{2}}{\alpha} + \norm{\bbeta^\star_{(\rho-\delta, \rho]}}_2.
\ee
The upper bounds \eqref{mybound} and \eqref{mybound2} hold as long as conditions \eqref{condition1} and \eqref{condition2} hold. 
By Lemmas \ref{lemma:reduced_model1} and \ref{lemma:reduced_model2} along with $N \geq c_0\, s \log p \geq \log p(0,\rho-\delta)$ ($c_0 > 1$) and a union bound,
the probability that \eqref{condition1} or \eqref{condition2} are violated is bounded above by
\be
\label{parameter.bound}
2\, \exp(- c_1\, N) + 6\, \exp(- c_2 \log p(0,\rho-\delta)).
\ee

Consider $\wm \geq \rho$.
Then the probability that 
\beno
\norm{\wh\bbeta - \bbeta^\star}_2
&\leq& \dfrac{16\, \sqrt{s}\, \lambda_{2}}{\alpha}
\ee
is violated is bounded above by \eqref{parameter.bound}.
\end{proof}
}

\hide{
\subsection{Proof of Theorem \ref{theorem.unknown}.3}

\begin{proof}
{\em Theorem \ref{theorem.unknown}.3.}
Consider $\wm < \rho$.
By (C.3),
for all $\delta > 0$ and all $\epsilon / 2 > 0$,
there exists $k_{\delta,\epsilon/2} > 0$ such that,
for all $k > k_{\delta,\epsilon/2}$,
the fraction of edges at distances $d \in [\rho-\delta, \rho]$ is at most $\epsilon / 2$.
Assume $k > k_{\delta,\epsilon/2}$.
Let $g$ be the fraction of false-negative edges.
There are two possible sources of error:
(a) some of the edges at distances $d \in [0, \rho-\delta]$ are not be detected by the two-step $\ell_1$-penalized least squares method,
and
(b) none of the edges at distances $d \in (\rho-\delta, \rho]$ is detected by design.
Let 
$g_1$ be the fraction of edges at distances $d \in [0, \rho-\delta]$ 
and
$g_2$ be the fraction of edges at distances $d \in (\rho-\delta, \rho]$.
Then
\be
\label{term1}
\mbP(g > \epsilon)
\hide{
\;=\; \mbP(g_1 + g_2 > \epsilon)
&=& \mbP(g_1 + g_2 > \epsilon)\s
\\
}
\;\leq\; \mbP\left(g_1 > \dfrac{\epsilon}{2}\right) + \mbP\left(g_2 > \dfrac{\epsilon}{2}\right).
\ee
{\bf (a) Term $\mbP(g_1 > \epsilon/2)$:}
Since $k > k_{\delta,\epsilon/2}$,
there exist at least $(1 - \epsilon/2)\, n > 0$ edges at distances $d \in [0, \rho - \delta]$,
where $n > 0$ is the total number of edges in the graph.
Denote by $\mG$ the event that all edges at distances $d \in [0, \rho-\delta]$ are detected and by $\mB$ its complement.
Then the event that the fraction of undetected edges at distances $d \in [0, \rho-\delta]$ exceeds $\epsilon / 2$ is contained in $\mB$ and the probability of the event of interest is bounded above by the probability of $\mB$.
To bound the probability of $\mB$,
observe that $\mG$ is implied by
\beno
\dfrac{2}{\beta_{\min}^\star}\, \norm{\wh\bbeta_{[0, \rho-\delta]} - \bbeta^\star_{[0, \rho-\delta]}}_\infty
&\leq& 1.
\ee
By \eqref{condition1} and \eqref{condition2} in combination with Lemma \ref{lemma.bound} and $\beta_{\min}^\star \geq 32\, \sqrt{s}\, \lambda_{2} / \alpha$,
\be
\label{ssss}
\dfrac{2}{\beta_{\min}^\star}\, \norm{\wh\bbeta_{[0, \rho-\delta]} - \bbeta^\star_{[0, \rho-\delta]}}_\infty
\leq \dfrac{2}{\beta_{\min}^\star}\, \norm{\wh\bbeta_{[0, \rho-\delta]} - \bbeta^\star_{[0, \rho-\delta]}}_2
\leq \dfrac{2}{\beta_{\min}^\star}\, \dfrac{16\, \sqrt{s}\, \lambda_{2}}{\alpha}
\leq 1.
\ee
The upper bound \eqref{ssss} holds as long as \eqref{condition1} and \eqref{condition2} hold.
By Lemmas \ref{lemma:reduced_model1} and \ref{lemma:reduced_model2} with $N \geq c_0\, s \log p \geq \log p(0,\rho-\delta)$ ($c_0 > 1$),
the probability that \eqref{condition1} or \eqref{condition2} are violated is bounded above by
\be
\label{source1}
\mbP\left(g_1 > \dfrac{\epsilon}{2}\right)\s
&\leq& 2\, \exp(- c_1\, N) + 6\, \exp(- c_2 \log p(0,\rho-\delta)).
\ee
{\bf (b) Term $\mbP(g_2 > \epsilon/2)$:}
Since $k > k_{\delta,\epsilon/2}$,
\be
\label{source2}
\mbP\left(g_2 > \dfrac{\epsilon}{2}\right)
&=& 0.
\ee
Combining \eqref{term1} with \eqref{source1} and \eqref{source2} gives \eqref{prob.b}.

Consider $\wm \geq \rho$.
An argument along the lines of (a) shows that, 
with at least probability \eqref{prob.b}, 
$g = 0$.
\end{proof}
}

\end{appendix}

}

\end{document}